\UseRawInputEncoding
\documentclass{ieeeaccess}
\usepackage{cite}
\usepackage{amsmath,amssymb,amsfonts}
\usepackage{algorithmic}
\usepackage{graphicx}
\usepackage{textcomp}
\usepackage{graphicx}
\usepackage{hyperref}
\usepackage[shortlabels]{enumitem}
\usepackage{tikz}
\usepackage{dsfont}
\usepackage{bm}
\usepackage{cases}
\usepackage{url}

\NewSpotColorSpace{PANTONE}
\AddSpotColor{PANTONE} {PANTONE3015C} {PANTONE\SpotSpace 3015\SpotSpace C} {1 0.3 0 0.2}
\SetPageColorSpace{PANTONE}%

\def\BibTeX{{\rm B\kern-.05em{\sc i\kern-.025em b}\kern-.08em
    T\kern-.1667em\lower.7ex\hbox{E}\kern-.125emX}}

\newcommand*\circled[1]{\tikz[baseline=(char.base)]{
    \node[shape=circle,draw,inner sep=0.2pt] (char) {#1};}}

\DeclareMathOperator*{\argmax}{arg\,max}

\makeatletter
\def\endthebibliography{%
  \def\@noitemerr{\@latex@warning{Empty `thebibliography' environment}}%
  \endlist
}
\makeatother

\begin{document}
\history{Date of publication xxxx 00, 0000, date of current version xxxx 00, 0000.}
\doi{10.1109/ACCESS.2017.DOI}

\title{On the Trust and Trust Modelling for the Future Fully-Connected Digital World: A Comprehensive Study}
\author{
\uppercase{Hannah Lim Jing Ting}, \IEEEmembership{Student Member, IEEE},
\uppercase{Xin Kang}, \IEEEmembership{Senior Member, IEEE},
\uppercase{Tieyan Li},
\uppercase{Haiguang Wang},
\uppercase{Cheng-Kang Chu}
}

\address{Digital Identity and Trustworthiness Lab, Huawei Singapore, North Buona Vista Drive \#13-01,  The Metropolis Tower 1, Singapore 138588}

\tfootnote{This work was supported by Shield Lab, Huawei Singapore.}

\markboth
{Author \headeretal: On the Trust and Trust Modelling for the Future Fully-Connected Digital World: A Comprehensive Study}
{Author \headeretal: On the Trust and Trust Modelling for the Future Fully-Connected Digital World: A Comprehensive Study}

\corresp{Corresponding author: Kang Xin (e-mail: kang.xin@huawei.com).}

\begin{abstract}
  With the fast development of digital technologies, we are running into a digital world. The relationship among people and the connections among things become more and more complex, and new challenges arise. To tackle these challenges, trust --- a soft security mechanism --- is considered as a promising technology. Thus, in this survey, we do a comprehensive study on the trust and trust modelling for the future digital world. We revisit the definitions and properties of trust, and analysis the trust theories and discuss their impact on digital trust modelling. We analyze the digital world and its corresponding environment where people, things, and infrastructure connect with each other. We detail the challenges that require trust in these digital scenarios. Under our analysis of trust and the digital world, we define different types of trust relationships and find out the factors that are needed to ensure a fully representative model. Next, to meet the challenges of digital trust modelling, comprehensive trust model evaluation criteria are proposed, and potential securities and privacy issues of trust modelling are analyzed. Finally, we provide a wide-ranging analysis of different methodologies, mathematical theories, and how they can be applied to trust modelling.
\end{abstract}


\titlepgskip=-15pt

\maketitle

\section{Introduction}
The world of digital data and information transfer is growing. Devices, once only capable of communicating within homogenous networks, can now transfer data between devices of varying background and capability. Technology continues to advance in this direction to allow devices, from sensors to smart phones, to communicate. As devices from different backgrounds connect, their physical and social environments become integrated with the digital world. Today, service providers --- that aid people, businesses and society --- increasingly utilise digital technology. Clearly, information exchange supported by the digital world has a large impact on society. Thus digital exchanges need to be safeguarded.

Integrating digital, social and physical worlds, however, exposes the digital world to newer and more complex vulnerabilities. Growth in the number and variety of entities in the digital world means digital exchanges are subjective and situational, with entities behaving and prioritising differently. It also means hard security, which provides widespread authenticated access control, is now unfeasible \cite{trustsecurity}. As mere participants, malicious entities can infiltrate and cause disturbances in digital networks.  Regulating behaviour to control the actions of entities and minimise their negative impact --- a softer form of security --- is now necessary to ensure digital communities remain conducive \cite{josangtrustmatters}.

Trust is a nuanced social concept instinctively used for interaction \cite{philobarber, philoluhmann}. Reflecting these social properties of trust into the digital world, allows digital entities to perceive others and choose their interactions, as is done in the real world. Therefore, trust, implemented as a soft security mechanism, provides much needed social management. Nevertheless, trust is hard to quantify; the perception mechanisms we use in the physical and social world are not available to implement in digital environments. Trust modelling is thus needed to mimic the evaluation and decision-making instinctively performed in real life. Efforts have been made in a number of digital environments and research continues to grow.

There have been a number of related trust survey papers. Shrikant and Sunilkumar performed a brief survey of trust models for Vehicular Ad hoc Networks (VANET) \cite{VANETsurvey}. A more comprehensive trust survey for a broader range of trust scenarios was proposed by Yan Zhen in her survey of trust for Internet of Things (IoT) \cite{iotsurvey}. Ruan surveyed trust in a different digital environment: online social communities.

In each of these surveys, the unique characteristics of the digital environment were discussed. In the VANET and IoT survey, the unique characteristics of the environment was used to outline key, wide-ranging, sometimes practical objectives for trust management. Yan Zhen went a step further, discussing how each of these objectives had to be addressed in each of their defined IoT layers. Ruan's online social community trust survey considered trust management objectives differently, recognising that there were attacks specifically targeting trust management systems, that needed to be addressed.

This analysis of environment for trust management is valuable for the implementation of soft security mechanisms for the real world. However, the digital world is broad and consists of many digital environments that differ, even when abstracted. Since each of these surveys only discussed one type of digital environment, it is not sufficient for a representative analysis of the digital world.

One of the key purposes of a survey paper is to breakdown existing literature. In Yan Zhen's paper, different models were classified based on their primary goal. Then, the models were evaluated based on whether they met the outlined trust objectives for IoT. In Ruan's survey for online social communities, different methods of understanding, computing and inferring trust was surveyed. Then, each trust model was evaluated based on its vulnerability to common attacks that undermine trust.

Categorising existing trust models into different methods and goals, and evaluating them is useful in determining the appropriateness of different types of models. Existing survey papers categorise the different models in meaningful ways and evaluate models from meaningful perspectives. However, in these papers, it is not always apparent whether it is the approach or the methods typically used in each approach that is insufficient. If the approach is insufficient, the approach should change. If it is the method that is insufficient, methods can be tweaked and improved upon.

There were some survey papers that took a attack and security-oriented approach towards trust modelling. In Wang's survey, different service-types were evaluated for their security requirements \cite{trustsecurity}. Different attacks and some models that addressed these attacks were discussed, though this discussion was relatively brief. A much more comprehensive attack and defense survey was done by Hoffman for reputation systems \cite{attackdefense}. In their survey, Hoffman presented a framework for decomposing reputation systems. In this system, the different system components and design choices that were vulnerable to attacks were discussed. Lastly, which defense mechanisms were most appropriate and how they could be incorporated into reputation systems for attack-resilience was discussed.

In both the above surveys, discussions about different security requirements is particularly valuable in understanding the extent of the trust research problem. The evaluations were also useful in giving an idea of appropriate model design choices for attack-resilience. However, in both survey papers, the range of attacks considered were limited. For a diverse environment such as the digital world, this range may be insufficient for implementation of secure trust management. Furthermore, more technical aspects of design choices were rarely discussed or evaluated. Therefore, these survey papers failed to offer a theoretical, technical baseline to expand on existing design choices to make them more suitable for defense. Instead, only the engineering, system-design direction was considered.

There were a few technical, method-oriented trust surveys. Guo classified different trust computation methods based on different design dimensions \cite{variousmethodssurvey}. They summarised the advantages and disadvantages of each dimension and highlighted whether they were effective against malicious attacks, particularly for IoT systems. A machine learning-oriented survey was carried out for trust management by Wang \cite{mlsurvey}. Covering different digital environments and rating methods, the machine learning methods that have been employed in different models were discussed. Each of the models were evaluated based on whether they could address the rubrics outlined.

Surveys that take the method-oriented approach are useful because they offer insight into suitable technical methods for trust management. However, Guo's survey was relatively brief and machine learning is often not suitable in many digital environments. Other technical methods may be more suitable but these were not discussed in Wang's survey on machine learning. A more detailed analysis of the theoretical and technical basis for trust modelling methods would be useful to the field. Furthermore, while methods are important, the factors considered are just as important. Many of these models did not consider, in depth, the appropriateness of factors and methods chosen to model specific factors. Hence, it is not known whether existing methods of modelling factors are appropriate and whether choice of factors is suitable for soft security.

The survey on computational trust and reputation by Diego discussed trust and reputation broadly from a computational, theoretical perspective \cite{modelsurvey}. In their survey Diego provided extensive definitions and concepts of trust and reputation. Then, they created a schematic to assess computational trust and reputation models. Finally, they analysed research directions taken by different models in the field. However, not all surveys can be examined via rubrics due to different standards and definitions in different models. Furthermore, the lack of mathematical analysis failed to offer insight into the best possible methods for individual digital environments. So, while Diego's survey is useful in examining state of the art in trust, it does not give much insight into the direction in which trust modelling should move. Furthermore, theoretical basis of models were rarely discussed which limits understanding of the most fitting methods for trust modelling.

In summary, the premise of many existing survey papers fail to address a broad enough scope for the digital world. Moreover, they fail to offer insights into appropriate factors, modelling and evaluation methods that will be suitable for the digital world. For methods, the gap in theoretical analysis is particularly stark. Hence, it is still hard to determine the best mathematical tools for trust. To address these gaps, our survey contributions are as follows:
\begin{enumerate}
  \item We analyse digital world environments and abstract them into the different types of trust that they require. Each type of trust has further interpretations that influence how trust is evaluated. To demonstrate our interpretation, we give examples from the digital world.
  \item Based on the different types of trust and using our understanding of each  digital environments, we propose sets of factors that can be used within the digital environment. These factors cover a very broad range of types to offer a holistic perspective on trust for better, well-rounded evaluation.
  \item We also use our understanding of digital environments to come up with a broad checklist for different trust model in different digital environments, along with trust-management attacks. Each model can be evaluated based on whether it can check of each of the boxes. A model that can pass the criterion is not only be secure, it is also usable.
  \item Finally, we look at different modelling methods and how they have been used. By offering technical details alongside relevant models, we offer insight into the usefulness and theoretical suitability for different digital environments.
\end{enumerate}
The rest of our survey paper will proceed as follows. In Section \ref{sec: trustchar}, we discuss key concepts relevant to trust and trust modelling, even discussing some social theories and how they relate to the digital world. In Section \ref{sec: digitalworld}, we outline the different digital environments, specifying the unique challenges they each face. Using our analysis of the digital world, we categorise trust into different types and provide factors that best model this type of trust in Section \ref{sec: trusttypes}. Next, in Section \ref{sec: methods}, we discuss different mathematical methods, the models that have used them and how they can be used in a general trust modelling framework. Lastly, we conclude and propose some future research directions in Section \ref{sec: futurework}.

\section{Trust Definitions and Properties} \label{sec: trustchar}
Trust is understood and used differently in different fields. In the humanities, trust and society have long been of great interest. In this section, we outline the most notable social theories of trust while exploring how they each inform digital trust. Then, we discuss digital trust, specifically its definition, properties and some crucial related concepts.

\subsection{Social Theories and the Digital World} \label{subsection: socialtheories}
Social theories about trust were pioneered by Simmel who contributed to the field in two ways. First, Simmel identified the function of trust, describing it as a force that works for and through human association, to bring society together \cite{simmelmoney,philosimmelhelp}. Second, Simmel explained its source, describing trust as a combination of inductive knowledge and faith \cite{philosimmelhelp}. Later, Luhmann expanded on Simmel's theoretical foundations. Luhmann explained that performing any action, no matter how basic, involved uncertainty and risk. Therefore, trust was necessary to assume at least the more unlikely risks were negligible, so individuals could function normally \cite{philoluhmann}.

Luhmann's and Simmel's ideas have been adapted by trust modelling. In the digital world, every exchange, no matter how basic, carries some form of risk. Like in the real world, trust is needed to simplify the large number of uncertainties so that necessary digital tasks can be performed. Digital trust uses evidence and implicit knowledge about the digital environment for reasoning and decision-making, like social trust, digital trust uses inductive knowledge and faith. Finally, when a digital environment implements trust, agents can utilise a basic social mechanism to interact. Trust is a synthetic force even in digital communities.

Later, Bernard Barber described how expectations form the foundation of interpersonal trust --- expectations that social mechanisms functioned properly and that others were willing and capable of fulfilling their roles \cite{philobarber}. This is highly relevant to the digital world. In the digital world, the ability and willingness to fulfill roles determines the success of interactions. Therefore, incorporating willingness and capability in trust modelling of digital agents would reflect the decision-making patterns from the real world.

More modern examinations of trust emphasise its ever-present necessity. Giddens discussed the emergence of social systems in the modern world and how these were founded on and helped sustained trust \cite{philogiddens}. Francis Fukuyama discussed the importance of trust for today's economic activities \cite{philofukuyama}. Most recently, Piotr Sztompka discussed trust from multiple perspectives \cite{philopiotr}. These discussions tell us that as we digitize more and more of our modern social and economic transactions, social interpretations of modern trust need to be reflected in its digital counterpart.

While modelling trust in the digital world may not directly depend on social theories, observations about society are still highly relevant to the network-like digital communities of today. Bearing social theories in mind allows more realistic trust modelling. Doing so, however, is challenging as the humanities only describe trust in its qualitative, vague and complex form. Practical applications of trust require the digital world translate conceptual trust into a tangible quantity. This is done in the next section by describing digital understandings towards trust.

\subsection{Digital Trust} \label{subsec: digitaltrust}
Digital agents exchange digital services and/or information, during which honesty and capability is required. Digital trust is defined as a "measurable belief and/or confidence" that is "accumulated from past experiences" and is an "expecting value for the future" \cite{itu-t}. To explain, this means trust quantifies one's certainty via sources of evidence, such as past experience. Evidence is accumulated and formulated into a prediction of future behaviour. In addition, there are a number of properties that need to be considered.
\begin{description}
  \item[Subjective] Trust levels differ between people \cite{itu-t}. Some digital entities are stricter, while others are more lax. Trust dispensed depends on each agent's own, potentially unobservable, preferences. Trust values and decision-boundary thresholds need to reflect individual preferences, not a universal standard.
  \item[Context-dependent]Trust levels also vary with context \cite{itu-t}. Different digital environments invite entities with different, sometimes malicious, intentions. Within each digital environment, each interaction differs from the other in truster, trustee, purpose and other, observable or unobservable, context features. Therefore, even within the most seemingly similar interactions, agents may behave differently.
  \item[Dynamic] Trust tends to wane with time \cite{itu-t,kangxin}. As time passes, any existing knowledge becomes increasingly outdated. So, expectations about an agent's future actions become increasingly uncertain. Without overcompensating, trust models need to sufficiently to reflect the decay of trust with time.
  \item[Transitive] Trust is transferrable \cite{itu-t}. When a trusted individual offers recommendations, the truster's preceding trust in the recommender implies a trust in the recommendation. The transitivity of trust drives the formulation of indirect trust which will be discussed in Section \ref{subsubsec: indirecttrust}. While trust may be transferrable, it should be noted that the extent of the transfer depends on the digital environment and individual agent.
  \item[Asymmetric] Trust formed between truster and trustee is directed,; the existence of trust in one direction does not imply existence in the other \cite{itu-t}. Therefore, a truster's belief in a trustee may not be returned to the same degree or even at all. In some cases, asymmetry can be very stark. Where there is an imbalance of authority, certified authorities are likely more trustworthy but will unlikely dispense trust as easily.
  \item[Easy to lose but hard to gain] In all digital environments, caution is exercised to some extent, to ensure security. This caution means agents tend to choose interacting with trusted, familiar agents over strangers and when trust is betrayed, it is usually forfeited relatively quickly \cite{kangxin}. While greater caution naturally implies greater security, it also implies fewer risks. So, some opportunities with trustworthy individuals are naturally forfeited. An appropriate balance is needed to ensure the digital environment is safe but functional.
  \item[Pervasive] Social theories from Section \ref{subsection: socialtheories} indicate that trust is an inherent prerequisite to any interaction. Its necessity makes it pervasive. As social communities move into the digital realm, this pervasive nature is observed in digital communities as well. Digital agents will find it impossible to interact and function in the digital world, without trusting other digital agents. So, like in the social world, trust is pervasive.
\end{description}

\begin{figure*}[t]
  \centering
  \includegraphics[width=\linewidth]{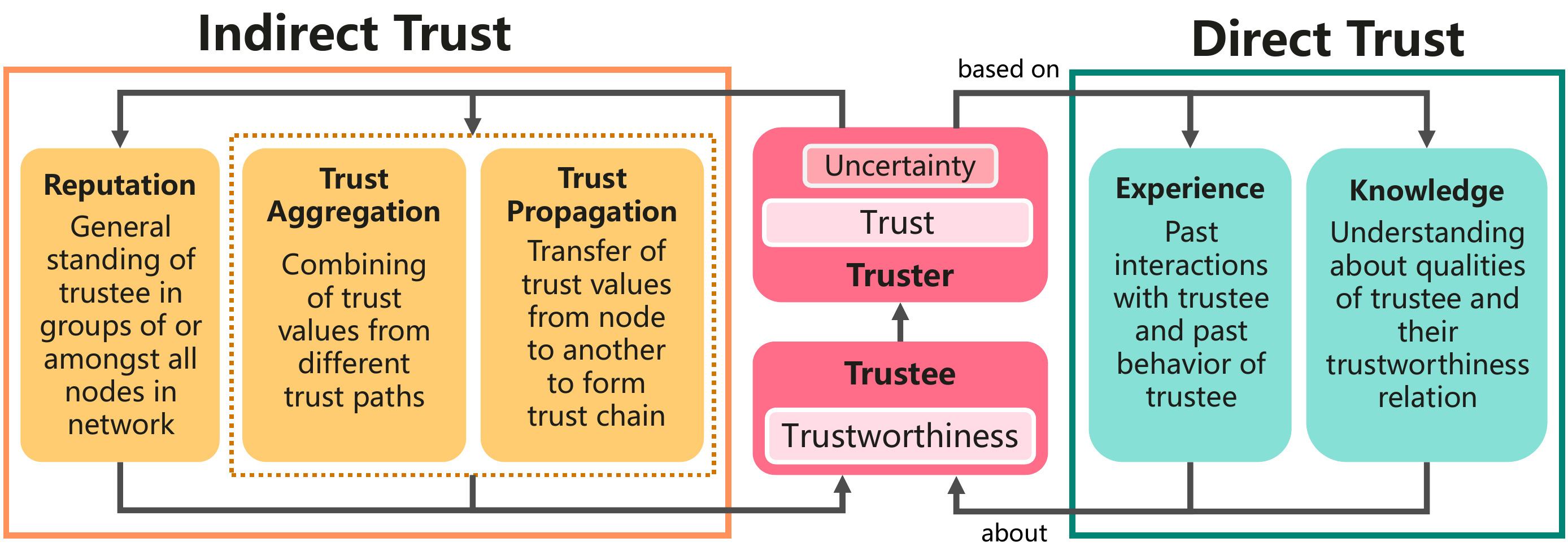}
  \caption{Flow chart of trust factors, concepts and agents.}
  \label{fig: trustconcepts}
\end{figure*}

In the next section, we define important terms and concepts. Figure \ref{fig: trustconcepts} gives a broad overview of these sub-concepts associated with trust. Broadly, trust in a trustee is broadly influenced by two factors: direct and indirect trust. Direct trust models personal knowledge about the trustee while indirect trust models the opinions of others. Both offer perspectives to form a more holistic view about the trustee. In the following sections, we will go into detail about each component.

\subsubsection{Trust Agents}
In trust management, the modelled entity determines \textit{what} is modelled. Typically, there are two entities of interest: the truster (the individual trusting) and the trustee (the individual being trusted) \cite{itu-t}.

Seen in Figure \ref{fig: trustconcepts}, a truster dispenses trust to a trustee, based on direct and indirect trust, about their trustworthiness. For similar evidence, different trusters may decide to trust differently. Trust propensity can be understood as the trustor's generalized expectation about the trustworthiness of trustees in general \cite{itu-t}. There are a number of influencing factors such as the level of security or degree of urgency. If an interaction was more important, for example when credit card information is being exchanged, higher trustworthiness would be required of the trustee. If a particular service is needed urgently, trusters may lower their expectations. Trust propensity illustrates that trust is subjective. Contextual features are necessary to capture this subjective. Direct and indirect trust will be discussed in Sections \ref{subsubsec: directtrust} and \ref{subsubsec: indirecttrust}.

Trust values assigned to trustees represent beliefs in the trustee's performance or behaviour. However, there may be uncertainty associated with the assigned value. This could be due to the reliability of evidence used; A trustee's past behaviour with the truster or with other recommenders does not guarantee their future behaviour. It is also possible that the amount of evidence is insufficient to be certain about the trust value of a trustee. A series of ten interactions provide greater certainty about trustworthiness than a single interaction. Ultimately, no method of modelling can fully capture trust. Trust is inherently subjective and vague. Representing trust quantitatively inevitably means missing some influencing variables. Trust values are therefore all inherently uncertain.

\subsubsection{Direct Trust} \label{subsubsec: directtrust}
Trust values derived solely from truster's personal opinions are represented by direct trust. Such personal opinions are formulated from past experiences with the trustee which give information about their intentions and capabilities thus giving insight into their future actions. However, behaviour can fluctuate, intentionally or otherwise. Such fluctuations need to be accounted for in a non-misleading way using other knowledge about the trustee \cite{nitti,vehicle20,DBNcontextaware, mokhtari, arefandtran, kangxin}.

\textit{Experience} includes past interactions and behaviour of a trustee \cite{chen2019,ZOLFAGHAR2011833,RSVM,Mohtashemi,frankwalter}. Past interactions refer to direct interactions that have occurred between the truster and trustee, and remove the need to rely on malicious recommenders. Typically, performance evaluation depends on interaction ratings, which are assumed to either be part of the environment or voluntarily provided by the truster \cite{frankwalter,zhang08,REGRET,FIREtrung}. Otherwise, binary successful and unsuccessful interactions are also used \cite{Mohtashemi,Che2015,hassanvector,kangxin, nitti}. However, ratings may not be available in certain digital environments. Each interactions is also contextually different and potentially irrelevant. Contextual features help elaborate on the nature of the interaction so that its relevance to present day can be determined and its contribution to direct trust adjusted.

\textit{Knowledge} could refer to trustee features such as their communities, capabilities and profiles. For example, the depth, detail and content of a user's profile could help reveal their authenticity and intentions \cite{chen2019}. Device features, such as computational capability, have also been factored into evaluating trustworthiness of devices in networks \cite{nitti}. Knowledge could also refer to contextual features. These include the nature and purpose of the interaction which indicate a trustee's incentive to perform well. Naturally, knowledge features need to be measurable and any proxy would encounter the same issue of uncertainty that trust values do. Nevertheless, trustee features are a useful tool to build a more holistic view of the trustee, their capabilities and intentions.

\subsubsection{Indirect Trust} \label{subsubsec: indirecttrust}
For a truster to interact with an unfamiliar trustee, direct interactions are insufficient to draw reliable conclusions. Moreover, knowledge about trustees are not always available or accurate. Indirect trust is an additional perspective to consider that instead, relies on the opinions of others.

A number of trust models perform aggregate and propagate trust values throughout digital networks \cite{KIMSONG11,STAR,frankwalter,zhang08,REGRET,FIREtrung, kangxin}. Between any two nodes, there may be one or more intermediate nodes in which a directed path can be formed, where each node provides a trust value for the node after it. This path can then become a chain of, presumably reliable, recommenders that results in input on the trustee, for the truster. This process of forming a trust chain is called \textit{trust propagation}. Given that direct connections are not always available, this method of gathering information from surrounding, trusted nodes becomes useful to patch any insufficient information. After propagating trust values, each path's trust values is consolidated. \textit{Trust aggregation} methods are needed here to consider which paths are trustworthy and to combine the different opinions.

Nevertheless, multiple opinions, numerous paths and the existence cycles make trust propagation and aggregation complex. When there are cycles or a large number of paths, it becomes difficult for computational methods, iterating through the network and all paths, to converge. How to combine and infer the opinions of other nodes also depends greatly on the application environment as recommenders may harbor ill-intention or unintentionally propagate inaccurate trust values. Some trust models have countered this by accounting for the amount of evidence \cite{zhang08,Che2015,DBNcontextaware} and reliability of the advisor \cite{zhang08,REGRET,nitti,FIREtrung, zeinab}. Path lengths are also a consideration as it is generally believed that the longer the path length, the more diluted the opinion \cite{KIMSONG11,STAR,ZOLFAGHAR2011833,maxflow,frankwalter}.

\textit{Reputation} is the general belief about a particular trustee. As in the social world, reputation can be derived globally, from all opinions \cite{zhang08} or from a select group of mutual acquaintances \cite{REGRET}. While methods of digital reputation may overlap with trust aggregation, rather than trust value inference, reputation is geared towards quantifying a node's surrounding structure and standing in a network --- the more qualitative aspects that influence general opinion. For example, reputation covers network features such as clusters and connection types. Clusters of direct connections could indicate some latent connections between nodes; the large number of outgoing to incoming edges could indicate that an entity is randomly forming connections. Therefore, reputation adds an important perspective to indirect trust.

\section{Digital Environments in the Digital World} \label{sec: digitalworld}
\begin{figure*}[t]
  \centering
  \includegraphics[width=\linewidth]{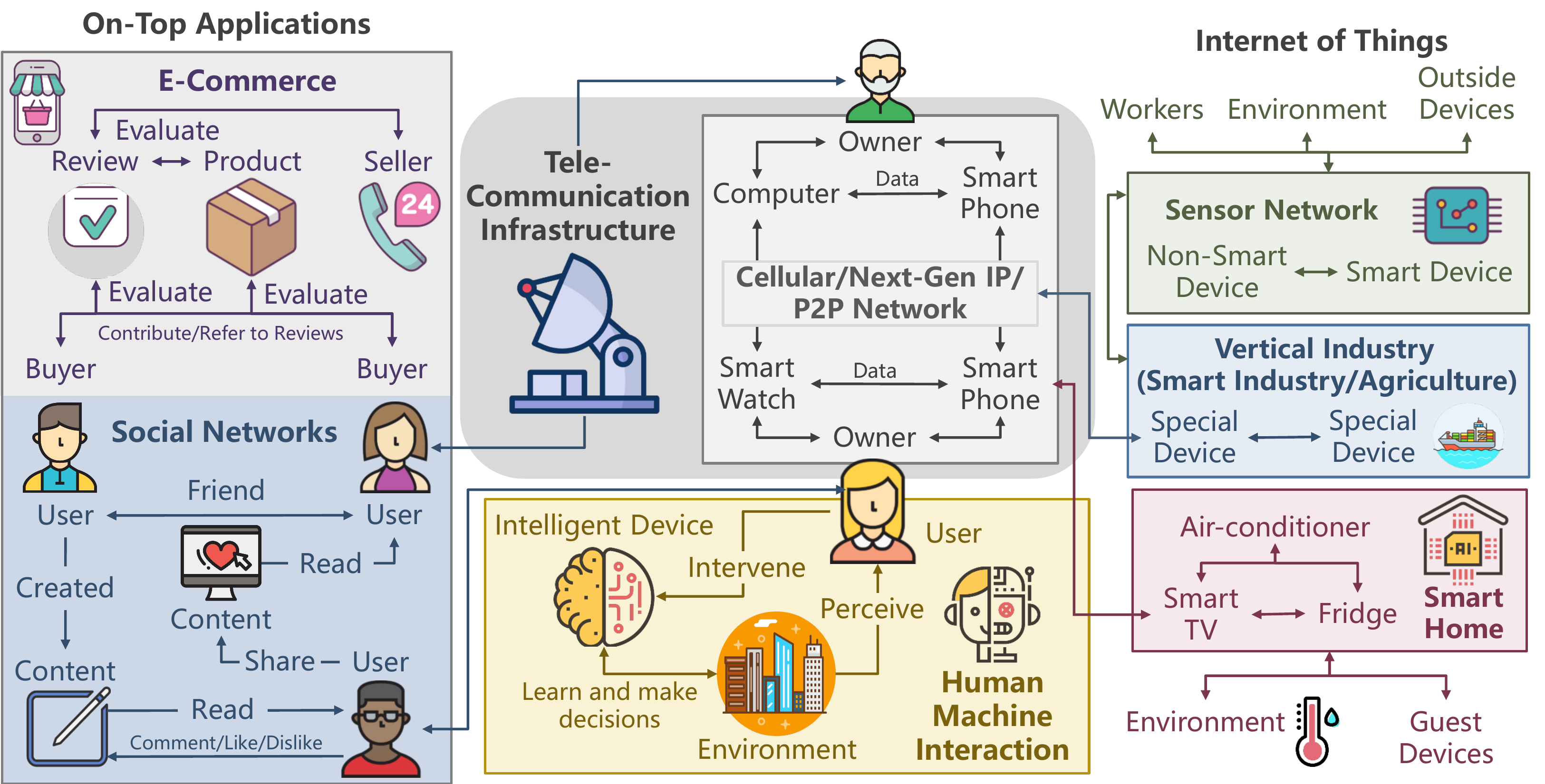}
  \caption{Outline of digital world applications and relationships}
  \label{fig: digitalworld}
\end{figure*}

There are many diverse environments in the digital world, each with unique characteristics and needs. Figure \ref{fig: digitalworld} illustrates the most representative digital environments and highlights some possible intra and inter-environment relationships.

There are many different types of relationships, each with unique features, that occur within digital applications and span across digital environments. Each digital environment may also support interaction with non-digital entities from the social world and physical environments. In this section, we use Figure \ref{fig: digitalworld} to discuss the different digital environments, their relationships and the respective trust and security challenges they face. This discussion will help inform the requirements and criteria for environment-appropriate trust management as well as the different types of trust, how to model them and what factors to consider.

\subsection{On-Top Applications} \label{subsec: socialecommerce}
On-Top Applications are upper layer type digital applications such as social networks and e-commerce platforms. In such free and open platforms, attackers can infiltrate and cause disturbances that render the network unconducive, driving users away \cite{josangtrustmatters}.

For example, in social networks, content created or distributed could be misinformation \cite{kbt,fakenewssurvey}, spam \cite{socialspam} or harassment \cite{homabullying}. Harassment is extremely detrimental to the mental health of its victims; When spam is rampant on a platform, users get plagued with large amounts of irrelevant information; The spread of misinformation on social platforms may cause users to act on misinformation with real-life repercussions. Harm generated in social networks tend to be widespread as other users can absorb and continue to propagate such harmful information.  Individuals need not even harbour malicious intent to help propagate damage. Trust is needed to manage interactions (sharing, liking, disliking, commenting and absorbing information) on social networks to reduce the impact of any harmful behaviour.

Harmful connections between digital users may also be formed. In some social networks, digital connections are able to access and receive updates about each other's content and personal information. Users may be granted privileges to directly converse with their connections. However, registration on most social networks do not require authentication so, agents can fake profile details. Malicious users can fake profiles to gain access to individual's private information or receive interacting privileges to harass or spread misinformation. It is then important for social networks to prevent these malicious relationships from being formed by determining which users are trustworthy and which are vulnerable.

There are similar content-trustworthiness concerns on e-commerce platforms, particularly with reviews. Reviews that are dishonest prevent sellers from earning off of their products and buyers from making good purchases. Such reviews could be left by competing consumers \cite{zeinab} or rival sellers to drive away competitors. While some reviews are clearly dishonest, there are more complicated contexts, with more grey areas. For example, reviewers may be disgruntled and or may not care about the quality of the review, resulting in partially true but potentially biased seller-ratings.

Reviews, however, are still important to prevent malicious and dishonest sellers from selling faulty or fake products. Therefore, reviewing, as well as buying and selling of products need to be protected by a trust mechanism. Trust measures are needed to disallow users from unfairly rating sellers or their products, without consequence. At the same time, reviews and ratings are also a trust mechanism to allow buyers and sellers to trust each other. In summary, trust management is useful for accurately reflecting product, seller and reviewer quality, for all users to make good decisions.

Privacy is of particular concern in on top applications. User data is distributed on a large scale in social networks and often collected on e-commerce platforms. Distribution and collection of personal information on a large scale puts the digital environment at risk of violating the privacy of users and potentially leaking confidential information. The design of trust management systems thus need to additionally consider safeguarding the transfer of large scale personal data \cite{itu-t}. Concurrently, trust management schemes frequently rely on identity and behavioural data of agents. Information used for trust management must not infringe on user privacy and maintain confidentiality, while also ensuring security.

With cross-application communication, the effects of malicious attacks on social networks or e-commerce environments can extend to affect the functioning of and users in other networks. For example, in Figure \ref{fig: digitalworld} a user in some social network could communicate false information to someone in a smart city environment, which then negatively impacts the functioning of smart cities.

\subsection{Cellular, Next-Generation IP and Peer-to-Peer Networks} \label{subsec: cellIPP2P_env}
Cellular, next-Generation IP and peer-to-peer Networks are some examples of communication networks supported by basic communication infrastructures. In such networks, communication may be less subjective and nuanced. However, other challenges exist. Devices vary greatly in capability and service flows differ between applications.

A single owner could have their devices communicate with each other. For example, in Figure \ref{fig: digitalworld}, a user could transfer video and audio files from their smart phone to their computer. During communication, each device needs to be sure that the data transferred has not been compromised. Not compromised means the owner of the device indeed has control, has authorised the transfer of information and the information is not, by some mistake of the owner, harmful.

Data from one owner's device can also be transferred to that of another owner. In such situations, the devices have a different relationship than co-owner devices. Data can be sent by malicious attackers to negatively impact some unsuspecting owner. Such data needs to be differentiated from intentionally transferred data. Even recognised devices could also erroneously or maliciously send corrupted files. Trust management can consider the variety of application scenarios and adjust trust values to different needs for differen situations.

In the above heterogenous networks, devices from a wide spectrum of capabilities are able to communicate. For example, in Figure \ref{fig: digitalworld}, data is transferred between a computer and a smart watch, a smart watch and a smart phone. Trust management is needed to prevent the spread of potentially compromised data, regardless of the capabilities of the weakest device --- the smart-watch. Trust management on networks need to be able to recognise malicious data and prevent its distribution, even if some devices involved are incapable of running large scale, computationally-heavy algorithms \cite{itu-t}.

\subsection{Internet of Things} \label{subsec: IoT}
Internet of Things (IoT) are networks of physical objects, integrated into information networks, to provide intelligent services. Physical objects include sensors, mobile devices and monitors. They extract information from their surrounding users and environment \cite{iotsurvey}. In Figure \ref{fig: digitalworld}, there are three primary digital environments: wireless sensor networks, vertical industries and smart home ecosystems.

In IoT digital environments, devices communicate within networks. In a smart-home environment, air-conditioners, refrigerators and other devices transfer data to facilitate intelligent home services. In vertical industries, special industrial equipment automatically communicate with each other to aid business operations. In wireless sensor networks, smart and non-smart devices communicate with each other to transfer collected data. Components within a network could be compromised when attackers launch malicious code to execute on IoT devices \cite{IoTattackssurvey} or when devices have been physically tampered with, intentionally by an attacker or unintentionally due to environmental conditions. Primitive devices may be incompatible with newer technology and malfunction.

IoT networks are particularly prone to such attacks and accidental errors. In outdoor vertical industries and wireless sensor networks, IoT devices communicate with numerous device types, the surroundings and personnel. When collecting data from the environment, its inherently complex nature makes it easy for devices to take erroneous readings. When devices communicate, data types and transfer modes may be incompatible and so data packets are lost. Furthermore, exposure to outside persons makes it possible that the devices could be physically moved or the data intercepted by malicious entities \cite{IoTsecuritysurvey}. Trust is needed to (a) identify when a IoT device within a network has been compromised and (b) when information transmitted between devices is erroneous or malicious, due to for example, spoofing attacks \cite{IoTattackssurvey}.

Devices can also communicate outside their network via telecommunication infrastructures. Data is periodically transmitted from IoT networks to user devices so that users can for example, monitor their smart home ecosystems. Any attacks on such IoT networks can have several repercussions. If data transfer to outside networks is not well protected, wormhole attacks could steal personal information and device passwords \cite{IoTattackssurvey}. When communicating with telecommunication infrastructure, denial of service attacks could disable IoT networks or prevent their access to larger networks for necessary services \cite{IoTattackssurvey}. Trust is needed for communication with outside devices to verify that devices will not steal information, reroute data packets or damage network functionality.

Challenges from managing different device types, mentioned in Section \ref{subsec: cellIPP2P_env}, are particularly evident in IoT \cite{itu-t, IoTsecuritysurvey}. Smarter, more capable devices consistently need to communicate with devices that are not primarily built to process data, such as sensors or refrigerators; Inter-environment communication means transmitted data tends to have largely different characteristics and requirements \cite{IoTsecuritysurvey}. Trust management needs to consider such data compatibility issues to determine if a device is trustworthy by, for example, considering the device's capability of fending off attacks. At the same time, management schemes also need to ensure low computational overhead for particularly resource constrained devices like those seen in IoT networks \cite{itu-t}.

Furthermore, the integration of inherently different physical, social and digital worlds poses unique security challenges \cite{itu-t}. Exposure to surroundings means IoT devices are more vulnerable to tampering, accidental and human errors. How IoT devices perceive information and whether data collected is reliable needs to be considered for trust. This means that besides malicious attackers compromising IoT and other digital devices, the presence of unreliable instructions from the social world and inaccurate data from the physical world also need to be considered.

One issue of great interest is the protection of personal data collected by IoT networks. In smart home ecosystems especially, IoT networks have gained entry to people's private homes and also receive data from personal devices and information networks. Personal information needs to be kept confidential and privacy preserved, especially because information is widely distributed to other IoT networks and telecommunication infrastructures \cite{IoTsecuritysurvey}. Simpler IoT devices, however, may not have the means to fend off against malicious interception of outgoing and incoming personal data. Entire IoT networks are then rendered vulnerable to privacy attacks. Trust management schemes therefore data flow. Simultaneously, trust management in IoT devices will likely require data about individuals, their profiles and their actions. Trust management schemes need to consider from which devices data is collected, how it is collected and how it is transferred. Excessive collection of information will invade user privacy and make potential leaks much more damaging.

Challenges in trust management for IoT networks are wide ranging. Besides identifying damaging data packets, trust for IoT needs to consider device constraints and differences in data type; data collection and distribution within IoT networks and the trust management need to be compliant with confidentiality and privacy standards. At the same time, for a trust system to be functional, data packets still need to reach their relevant destinations to ensure the functioning of these IoT digital environments \cite{iotsurvey}.

\subsection{Human Machine Trust}
Human-machine collaboration, where people interact with artificial intelligence and smart devices, is an emerging digital environment in trust. It has garnered interest for its usefulness in assisting decision-making and automating tasks and has already been employed in medicine, education, industry and space exploration \cite{HRCsurvey}. In human-machine collaboration, artificial intelligence in smart devices interact with people and the environment to take or propose appropriate actions \cite{huhmi, crimepredictionhmi}.

However, artificial intelligence is not always accurate and may make errors. Due to these errors, users may be tempted to distrust the machine and constantly override it, rendering the machine useless \cite{huhmi}. For a person to rely on machines and artificial intelligence, trust is needed. Trust allows people using machines to know when and under which conditions are machines trustworthy. This then allows the benefits from incorporating them into industries and service provision to be realised.

\section{Trust Types for the Digital World} \label{sec: trusttypes}
\begin{figure*}[t]
  \centering
  \includegraphics[width = \linewidth]{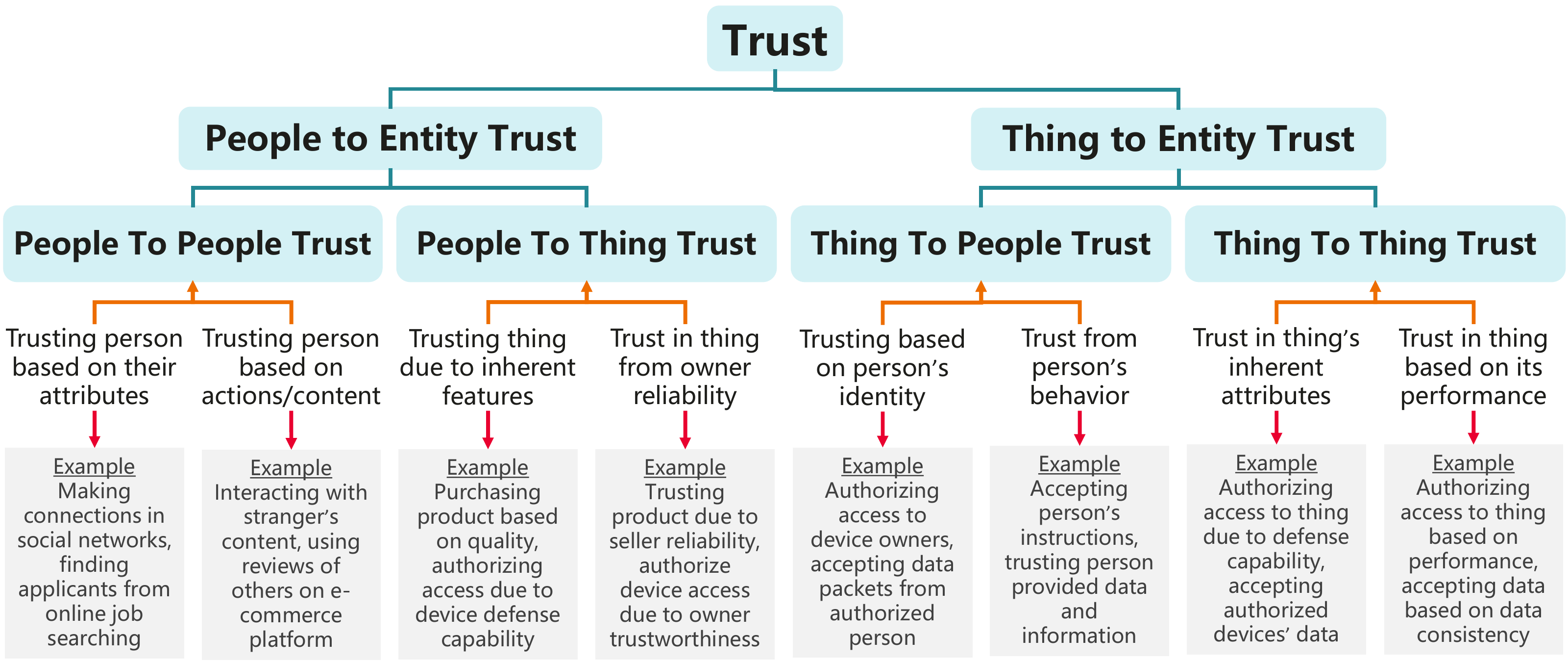}
  \caption{Organizational chart of different types of trust, how they are observed and examples.}
  \label{fig: trusttypes}
\end{figure*}

In Section \ref{sec: digitalworld}, digital environments, their need for trust and each of the unique sets of challenges they faced were discussed. Now, we consider how to formulate trust conceptually to best target each environment. By categorising trust into different types, we consider different scenarios, agent relationships and propose relevant factors. Figure \ref{fig: trusttypes} summarises the different types of trust with examples of their application scenarios and digital relationships.

\subsection{People to People Trust} \label{subsec: peopletopeople}
\begin{figure*}[t]
  \centering
  \includegraphics[width=\textwidth]{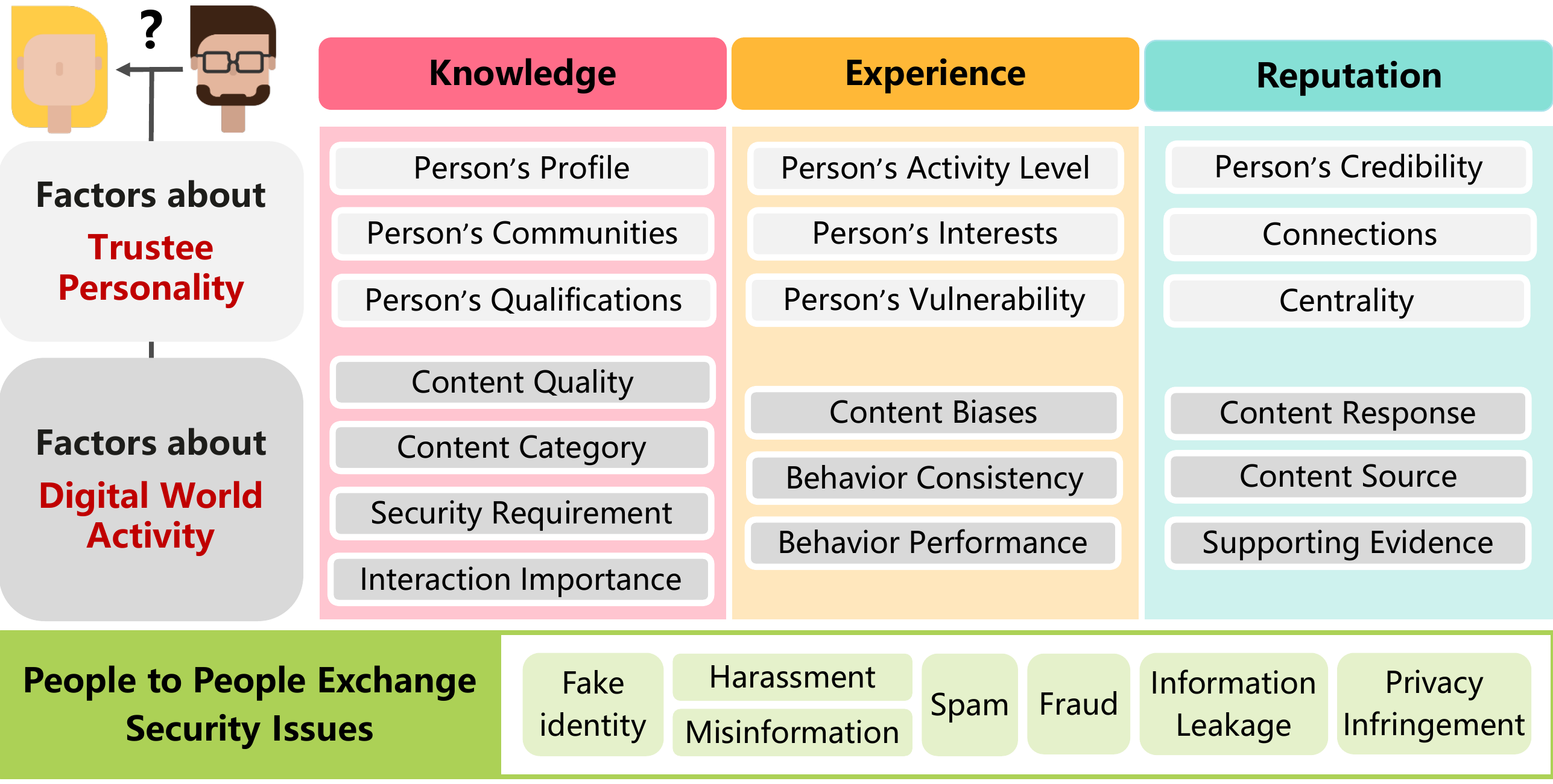}
  \caption{Overview of people to people trust factors and security issues.}
  \label{fig: peopletopeople}
\end{figure*}

People to people trust is formed between any two people with some relationship. In such relationships, people may fake their identities, harass others, spread misinformation, spam others, commit fraud, leak information or infringe on others privacy. With reference to Figure \ref{fig: trusttypes}, to verify that trustees will not commit any of these acts, trusters need to \textit{trust that a person is inherently good} or \textit{trust that a person will not do harm}. This interpretation helps formulate factors that influence trust between people. They are outlined in Figure \ref{fig: peopletopeople} along with the relevant security issues.

\subsubsection{Trust in Person's Inherent Goodness} \label{subsubsec: peopletrustformation}
People can trust that a person is inherently good. In this kind of person-centric trust, a truster believes that the identity and features a person presents themselves under is truthful and that they are inherently honest. In Figure \ref{fig: digitalworld} and Section \ref{subsec: socialecommerce}, we described how users makes "friends". To befriend others, users on online social networks need to trust the authenticity and features of a profile. On online job advertising platforms, employers and potential hires connect based on user profiles, features and identities. In such environments, trusters utilise general understandings about trustees to formulate a perception about them which forms their trust.

Knowledge about a trustee's traits and features describes them and who they are. This perception of trustees, combined with generalised expectations of people with similar traits, help trusters understand whether the truster is inherently trustworthy.
\begin{description}
  \item[Person's Profile] On social networks, user profiles are widely available and can contain information such as a user's age, location and write-ups about themselves. The quality of a profile can be very telling. Spam accounts lack incentive to invest much effort into profile creation so, the apparent amount of effort helps differentiate such accounts \cite{chen2019}. Profiles with many inconsistencies, suggest users are lying and dishonest. Profile features also partially describe trustees. Understanding trustees personality and abilities gives insight into their trustworthiness.
  \item[Person's Communities] Another way to describe a trustee is to use their communities and interests. Similar users are more likely to consider each other trustworthy \cite{ZOLFAGHAR2011833,DSNN}. Knowing a trustee's communities and interests tells us which topics trustees are invested in, allowing them to be paired with more like-minded trusters. For example, on social networks, if two users belong to the same networking group, they are more likely to share the same opinion. Therefore, they are more likely to find each other trustworthy.
  \item[Person's Qualifications] In some cases, qualifications are relevant. A person's qualifications determines if a suitable connection can be formed for hiring on online job markets. On e-commerce sites, a reviewer can be particularly qualified to recommend certain products. With these qualifications, users directly perceive trustworthiness based on inherent characteristics. It should be noted that qualifications can be inherited from the real world but they can also be implemented by trust systems based on environment-specific metrics.
\end{description}

To understand who a person is, trusters can also rely on past experiences with trustees. With experiences, trusters can gather evidence about a trustee's personality and reliability. This understanding of the trustee's personality directly determines their trustworthiness.
\begin{description}
  \item[Person's Activity Level] Activity levels indicate the level of investment in a digital community. Trustees that participate less, are less invested. A lack of emotional investment suggests users care less, so their activity could be more error-prone. Activity patterns in different communities also inform which interests trustees are invested in. Generalized expectations about people with certain interests form perceptions of trustees to different trusters.
  \item[Person's Interests] Another way to describe personality is through interests. The topics that trustees respond to are an indicator of their level of interest in different topics. User interest can differentiate between trustworthy and untrustworthy trustees for specific trusters. For example, if users have a similar pattern of reviewing products on online markets or respond similarly to content on social networking sites, they likely have similar interests and so will share similar opinions \cite{DSNN}.
  \item[Person's Vulnerability] Trustees that are more exposed to or more trusting of damaging content are considered vulnerable. More vulnerable trustees have a higher likelihood of distributing and spreading harmful content. So, content passing through or originating from vulnerable nodes should be considered less trustworthy. At the same time, reducing a gullible node's trust propensity when they are trusters, reduces their exposure to damaging content. As trustees, they are then less vulnerable, thus they are more trustworthy. To measure vulnerability, the degree of absorbtion of malicious content can be measured, normalized or multiplied by a trustee's exposure.
\end{description}

Reputation factors quantify the opinions that other users have towards the trustee. With the opinions of others, trusters can gain a better insight into the trustee's personality, particularly their behaviour with other people other than the truster themselves. This patches any information the truster does not know about a trustee's personality.
\begin{description}
  \item[Person's Credibility] A user that is credible is a user with a good global reputation. Being credible can be measured by the structure and distributions of trust values surrounding a node. While specific methods to measure credibility may coincide with trust aggregation \cite{KIMSONG11,STAR,ZOLFAGHAR2011833,DSNN,Mohtashemi,maxflow,frankwalter}, global network-based indicators such as popularity or authority score, adapted from PageRank, have also been used \cite{chen2019,ZOLFAGHAR2011833,info-trust,DSNN}. Rather than mere aggregation, PageRank-inspired popularity indicators consider the number and quality of incoming edges to measure credibility. Credibility thus ranks a user in relation to the entire network.
  \item[Connections] Connections measure reputation using a similar underlying concept as credibility but restricted more locally. Connections of trustees and trusters are both of interest here. Connections a truster has is important because they form a group of people the truster believes. Therefore, a truster's community's opinion about a trustee would be important in helping a truster evaluate trustworthiness. A trustee's connections are also important because beliefs about a trustee's community are generalized to the trustee themselves.
  \item[Centrality] Between a truster and a trustee, centrality is the degree to which a truster's network is central to the trustee or vice versa. For certain digital environments, if a truster's network is central to a trustee's, this suggests the trustee is important to the truster therefore, the trustee is more likely to behave favourably as their attention is less divided. Vice versa, if a trustee's network is central to the truster's, this could, in some digital environments, suggest that the trustee is already well-connected with people the truster is familiar with. Therefore, the trustee is already quite reputable amongst the truster's circle of friends.
\end{description}

\subsubsection{Trust in Person's Actions}
Trusting that a person is inherently good may be too na\"{i}ve in some cases. Sometimes, trust can only extend to believing that a person's current and future actions are not harmful. We have described content-type interactions between people in Figure \ref{fig: digitalworld} and Section \ref{subsec: socialecommerce}. On e-commerce platforms, users can evaluate reviews and determine that reviewers that have no incentive to leave harmful reviews. In social networking, users evaluate content to determine that content creators are objective. In both these cases, trusters interact with trustees because they believe trustees and their behaviour is not harmful.

To determine if a trustee is behaving in a non-harmful, objective and fair manner, we require descriptive factors about a trustees actions and how these related to a trustee's future trustworthiness. They are knowledge-type factors that describe a trustee's actions. In addition, we can also consider knowledge about trusters and their capacity to trust the actions of trustees.
\begin{description}
  \item[Content Quality] A trustworthy user's content reflects their quality. Quality can mean a number of things. First, it could mean factual accuracy. Comparing presented facts with known facts and counting the number of inaccuracies, we can determine if a user is objectively wrong. This method, however, na\"{\i}vely assumes no grey areas. Second, quality could refer to presentation. In word-heavy digital communities, grammar, choice of words and punctuation reveal the amount of effort invested and emotions in a piece of writing. An excessively emotional piece could be biased and writing riddled with grammatical errors and vulgarities are presentation-wise, similar to harmful content like spam and harassment. In this sense, quality of presentation indicates content trustworthiness.
  \item[Content Category] People's taste and preferences are indicative of their biases. The categories of content they typically engage with is indicative of this taste and preference. Understanding these potential biases, helps form an expectation of the trustee. We can then infer their potential biases in future related content. Then, without having to deterministically process any information, trusters can exercise caution with certain trustees about specific topics.
  \item[Security requirement] Determining the trustworthiness of content on online social communities need not rely only on the specific trustee's activity. Online social communities can differ greatly in policy and purpose, thereby attracting different kinds of people. For example, some online communities have stronger anonymity policies, reducing accountability and allowing users to be more irresponsible. Therefore, certain communities may see higher rates of malicious activity. Users in such communities need to exercise greater caution so, trust values should be lowered for tighter security.
  \item[Interaction importance] Different users use digital services for different reasons. Some users may perform important transactions that have a large potential loss. For example, a user that is buying a big-ticket item on an online marketplace should be more weary of dubious sellers and dishonest recommendations than if they were buying something much cheaper. During such interactions, users need to be less trusting, meaning that trust values need to be lowered to fit the requirements of each interaction.
\end{description}

To evaluate if a trustee will or will not do harm, we can look at their past behaviour as an indication of their tendency to cause harm. A trustee that has been harmful in the past, has shown that they cannot be trusted to behave well in the future. To evaluate their possible future harmful actions, we discuss experience type factors.
\begin{description}
  \item[Content Biases] In online social communities, users lack incentive to be objective. For example, e-commerce reviewers who have had bad experiences may be inclined to give an exceptionally bad score, even if their experience was not objectively as bad. The pattern of behaviour, choice of words and manner of writing on online digital communities give an indication if a user tends to favour certain positions \cite{DSNN}. If a trustee tends to overreact, they are more likely biased so, they are likely less trustworthy.
  \item[Behaviour Consistency] Consistency in a person's content reflects whether they behave in a consistent manner and thus, whether or not they can be trusted over time. A user that is more consistently good, provides a larger proportion of positive evidence, meaning their future behaviour is more certainly reliable. There are no instances of negative behaviour that may breed uncertainty in positive judgements. Alternatively, consistency can refer to informational consistency. If a user tends to contradict themselves within or between activities, this indicates the user is logically inconsistent, so they are untrustworthy.
  \item[Behaviour Performance] While consistency determines if evidence about a trustworthy can give a certain outcome, the performance of a trustee determines what the outcome is. Obviously, if a trustee performs well, this performance supports that they are trustworthy. Otherwise, a user is untrustworthy. The magnitude of performance is particularly telling in people to people trust. Good content requires more time and effort. A more invested trustee is more trustworthy.
\end{description}

Finally, we use the opinions of others towards the action of a trustee to determine if the trustee's actions are harmful. By looking at how reputable the actions of a trustee is, we can determine if a trustee is likely to do be harmful. These are reputation factors.
\begin{description}
  \item[Content Response] Response to content is made up of reaction mechanisms such as liking, disliking, commenting and the pattern of propagation throughout the network. Such a method of evaluating content is beneficial as it borrows from the opinions of real people, who are better able to perceive nuances. As individuals respond, the manner of distribution, who the content is propagated to and who propagates the content, is recorded. Based on the similarity of distribution patterns, the nature of content can be inferred. By proxy, we can determine if the creator is trustworthy.
  \item[Content Source] When content is distributed, the original source of information heavily implies the content's trustworthiness. If the originator of information is generally untrustworthy, it becomes highly likely the content is untrustworthy. That the trustee propagated untrustworthy information also reflects poorly on them.
  \item[Supporting Evidence] Typically, to evaluate the factual accuracy of some piece of information, people will compare the information to sources they consider trustworthy. In trust, content can be directly compared to other sources. Ideally, these sources should be formal or highly-regarded. Other sources of information act as supporting or disproving testimonies. The more independent, trustworthy supporting testimonies there are, the more likely content is trustworthy.
\end{description}

\subsection{People to Thing Trust} \label{subsec: peopletothing}
\begin{figure*}[t]
  \centering
  \includegraphics[width=\textwidth]{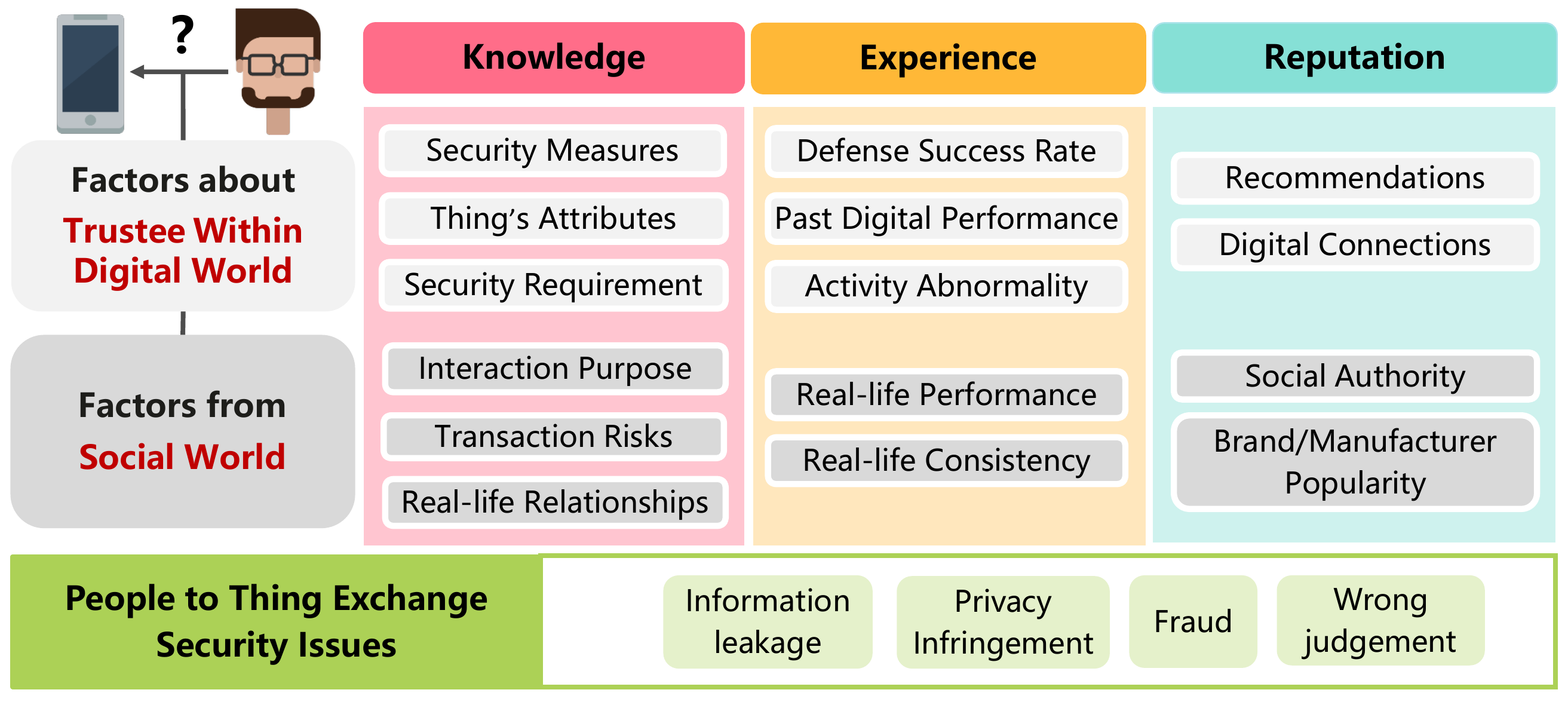}
  \caption{Overview of people to thing trust factors and security issues.}
  \label{fig: peopletothing}
\end{figure*}

People to thing trust is established from people to non-sentient entities, such as devices, products or information. People to thing trust crosses the divide between social and digital worlds; People trusters use social trust mechanisms to perceive trustee objects using only their characteristics presented within the digital world. In this dual-world, devices can leak information or infringe on the privacy of people; products can be fraudulent; devices can provide wrong judgements in human-machine interactions.  To trust things, people can use their \textit{digital world perception of the object} or \textit{their real, social world perception associated to the object} to evaluate if the object is trustworthy. We use this understanding to propose some factors for modelling. These are outlined in Figure \ref{fig: peopletothing}.

\subsubsection{Trust based on Digital World Perception}
Using their perception of the object in the digital environment, trusters can determine if it is of good quality. In Figure \ref{fig: digitalworld}, we illustrated that buyers in e-commerce markets need to evaluate product trustworthiness. To do so, buyers evaluate the product based on its digital world characteristics. They may consider product features, brand, build and materials to determine product quality; Users may use price to determine if a product is "too good to be true" or not worth the price, making purchase decisions based on cost. In peer-to-peer networks or smart home ecosystems, people interact with devices. Users may allow devices with strong defense capability to determine whether or not to accept incoming data. These demonstrate that people can trust in things using only the object's digital characteristics.

Knowledge factors can tell us if an object in the digital world possess characteristics that make them trustworthy. This is also influenced by other characteristics of the digital world that may influence people's perception of the digital object.
\begin{description}
  \item[Security measures] Security measures refer to mechanisms deployed to fight malicious attacks. Measures whether deployed by trusters or trustees are both important. Trusters with capable security mechanisms can exercise less caution. In cellular, next-Generation IP and peer-to-peer networks (discussed in Section \ref{subsec: cellIPP2P_env}), devices that have anti-virus software installed can trust other devices more readily. In e-commerce networks, a buyer can readily purchase products if they have buyer's insurance policies. Trustee's security measures are also important. Trustees that are not well-protected lack capability; capability in the sense that they are less able to provide non-harmful services to trusters with high success rate. For example, a device with no anti-virus software is more likely to transfer compromised files. Trustee devices should be less inclined to receive files from poorly-protected devices.
  \item[Thing's attributes] Obviously, thing features are important to determine trustworthiness. Trustworthiness entails that the thing can function as needed, without causing harm. Features considered differ depending on environment. In people to device environments, we can consider a device's build, brand and computational capability to determine if they are truly capable of performing necessary tasks, without corrupting files. Whether a product is trustworthy, is determined by their quality. In people to product digital environments, the amount of information available about the product, product-specific specifications, pictures and materials give indication of authenticity and reliability.
  \item[Security requirement] Since people to thing interactions can have very real implications, users generally have scenario-specific requirements to ensure an ideal set of outcomes for themselves. These requirements determine trust propensity. If users are engaging in a high-risk or high-cost activity, like exchanging large amounts of personal information or buying an expensive product, they are likely more cautions and less trusting. In scenarios where digital environments are ridden with dishonest users, trusters should trust fewer trustees to reduce their likelihood of encountering malicious agents. In this case, security requirements are higher so, trust propensity is lower.
\end{description}

Experience about digital world activity tends to apply more to device behaviour. Understanding the past behaviour of a device helps formulate a person's perception of how trustworthy and reliable a device is in the digital world. Digital devices that generally do not perform well are likely old, too primitive or compromised.
\begin{description}
  \item[Defense success rate] Devices with good defense success rate is resistant to malicious attacks. Being resistant means it is unlikely the device has been compromised before. Therefore, when interacting with the device, users can be more confident there is no existing malicious software in the device that can steal information or cause harm. A good defense rate is also evidence that a device can deliver under malicious threat. Therefore, when interacting, users can be more confident the interaction will be successful even if there were any attempts at a malicious attack on the device.
  \item[Past digital performance] Past digital performance of a device gives an indication of a device's ability to perform tasks and any potential malicious intent. Devices with good past digital performance means they have the computational capability to handle their assigned tasks, thus far. That devices fully provide necessary services also implies they have not been compromised by malicious attackers to cause disturbances in the network. Therefore, good digital performance is indication the device is trustworthy.
  \item[Activity abnormality] Finally, without external interception, devices should typically function as per normal. Abnormal behaviour in devices is indication that devices have been compromised. Compromised devices are more likely to cause harm and fail at providing services. Therefore, they should be considered less trustworthy until they are fixed. Abnormal behaviour can be measured by deviation in the devices behaviour from what is normal. For example, if a owner notices unusual power consumption, this could mean the existence of additional malicious background software, especially if nothing was done to trigger such power consumption.
\end{description}

Lastly, we can use the opinions of other devices to adjust our perception of a object as a digital world entity. These opinions tell us which devices regard the device in what way. Based on other device perception, people can determine if a device is likely to have been compromised.
\begin{description}
  \item[Recommendations] Recommendations are useful in people to thing digital environments where users are not always familiar with all the devices and products they interact with. Having recommendations, users are better able to determine if a product or device is trustworthy based on the opinions of others. For example, in e-commerce markets, users typically do not purchase the same product multiple times. Recommendations by other users about products then help users make better decisions about which products are more reliable. This has been discussed in detail in Section \ref{subsec: peopletopeople}.
  \item[Digital connections] Devices and products can possess connections to other devices and products. These connections allow one to infer about the trustworthiness of the target trustee. For example, if a product is associated --- by seller, brand, manufacturer or any other feature --- to a disreputable product, buyers should be less inclined to buy the target product less they possess they same issues as their connections. A device that has connected with compromised devices have been exposed to malicious or damaged devices. Therefore, there is a higher chance the target trustee device has been compromised. Again, users should be less inclined to trust target trustees with disreputable connections.
\end{description}

\subsubsection{Trust in Thing Based on Social World Relation}
It is not always possible to evaluate an object based purely on its characteristics. However, there are often people intermediaries between trustees (things) and trusters (people). Trust in a thing can be inherited from existing trust in people. Examples are illustrated in Figure \ref{fig: digitalworld}. On e-commerce platforms, the quality of a product cannot be fully verified. However, users can still choose to buy products anyway because a seller can be trusted to deliver authentic, true-to-picture and good quality products. In the case of smart homes, guest devices can connect to home networks if the guest device owner and home owner know each other. In both these cases, people to thing trust can is established because trusters have some form of guarantee. In the case of smart homes, this guarantee is not observable in the digital world.

Social world relationships influence people's decisions. However, social variables are generally latent from the digital perspective. We can use knowledge factors to understand device owners and their social world relationships.
\begin{description}
  \item[Interaction purpose] Interactions are generally undertaken with some goal in mind. Interaction goals aid trust modelling in two ways. A trustee's goal when participating in interactions tell us how motivated they are to perform well. If users stand to benefit from an interaction, they are more likely to cooperate. For example, trustee device owners have more incentive to behave cooperatively if the interactions will significantly up the reputation score of the owner. Therefore, the probability of success increases. Second, purpose of interaction informs the level of access trustees should have access to. For example, to provide services, devices typically only require access to some, and not all, personal information. If devices request for say, full access, this should arouse suspicions. Otherwise, understanding the purpose of any interaction helps tailor the trust threshold for decision-making.
  \item[Transaction risk] Some interactions carry more risk than others. In high-risk transactions, trusters should generally exercise more caution. For example, when purchasing expensive products online, trusters should exercise more caution in ensuring the quality of the product and the seller honesty before putting money down. Trust evaluation needs to be more stringent in high risk cases to reduce the odds of large losses.
  \item[Real-life relationships] Since social world relationships heavily influence people's trust in related devices, the existence and nature of real-life relationships naturally influence whether a person truster trusts a device owner. For example, when a device requests to connect to the network in a person's smart home ecosystem, it is generally ill-advised to simply accept a stranger device's request. However, if the two owners know each other, even if digitally, the truster is a stranger to the device, real-life relationships between the device owner and truster allows a trust relationship to be formed. Such real-life relationships should be considered wherever possible as they are influencing variables that cannot be observed in the digital world.
\end{description}

Like in people to people trust, experiences with social intermediaries is evidence of their trustworthiness. This trustworthiness is inherited by the associated thing as good experience with the social intermediary offers guarantee when interacting with the associated thing.
\begin{description}
  \item[Real-life Performance] Based on past experiences with an associated person, trustee things can inherit trustworthiness from their associated social world agents. Being confident in associated entities offers greater guarantee for success when interacting with the thing. For example, in e-commerce networks, trusters may be familiar with a particular seller, knowing the seller often sells high-quality and authentic products. They can use this familiarity to evaluate product reliability, even if it cannot be directly verified.
  \item[Real-life Consistency] Consistency is relevant to experiences with intermediaries. If a trustee intermediary exhibits good but inconsistent performance, there is a greater level of uncertainty associated with their trustworthiness. Therefore, less trust should be assigned to the intermediary so, less trust is passed on to the object. This can be seen in online marketplaces. If a seller is fairly inconsistent in delivering good products, trusters will be more cautious about purchasing from them, less the truster is unlucky and the seller fails to perform. Decisions to trust the product by buying it are then less likely.
\end{description}

Finally, products and devices are made in the context of the real, social world. Within the social world, things have attributes with social reputations tied to them. Using these related reputation factors, thing trustworthiness can be determined. We discuss them here.
\begin{description}
  \item[Social authority] If an object's intermediary has a good reputation, their increased social authority implies they are more trustworthy. This trustworthiness is inherited by the truster. For example, if a device owner is a verified government personnel, by formal standards, it is likely devices deployed by such personnel are more trustworthy. Therefore, even if the device is new to the network, they inherit social authority to become more powerful within the digital network.
  \item[Brand/Manufacturer popularity] Trustee things may also possess attributes related to the social world. Trusters use their generalized expectations to form an understanding of the thing based on its attributes. For devices or products, the most relevant attribute would be brand or manufacturing popularity. People generally form expectations about brands and manufacturers based on hearsay and their own experiences. If a brand is known for producing high quality goods, so long as the goods are truly from the seller, trusters are more likely to purchase the branded goods as quality is, to the truster, guaranteed. If devices produced by certain manufacturers are known for being secure, users may use this understanding to choose devices by such manufacturers. Associated real world understandings of intermediaries thus influence object trustworthiness.
\end{description}

\subsection{Thing to People Trust}
\begin{figure*}[t]
  \centering
  \includegraphics[width=\textwidth]{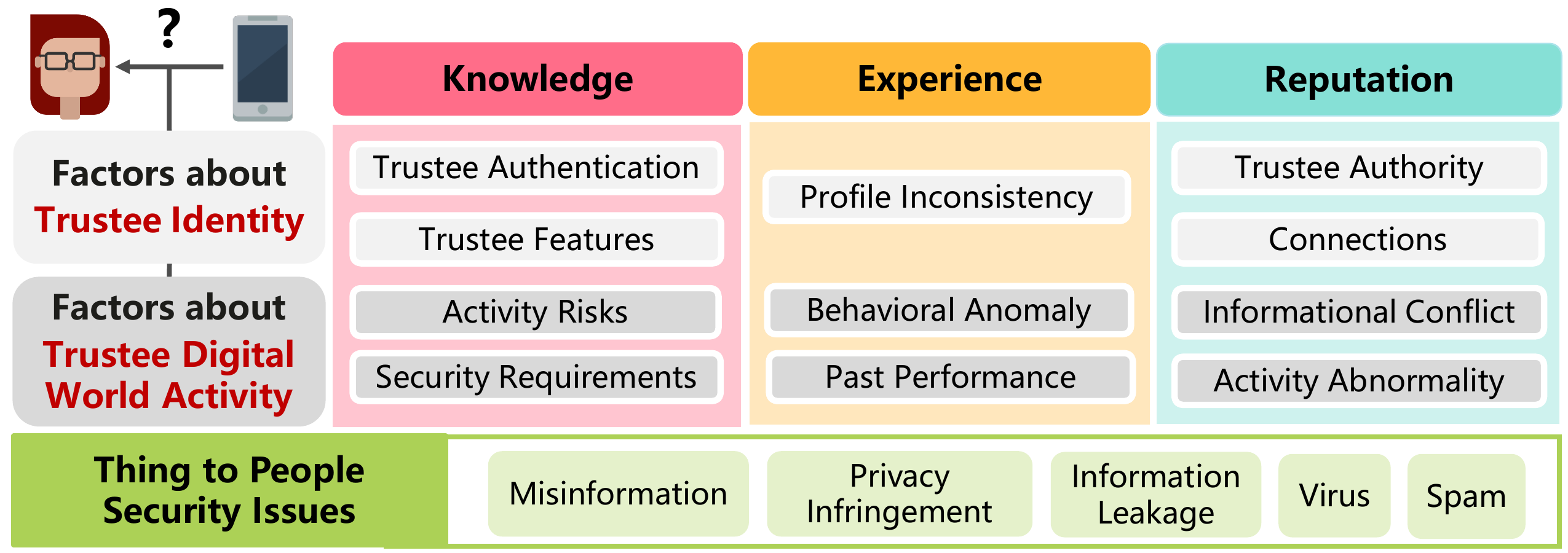}
  \caption{Overview of thing to people trust factors and security issues.}
  \label{fig: thingtopeople}
\end{figure*}

In an increasingly digital society, people become agents in device networks. Devices receive information --- instructions or data --- from people and have to make trustworthiness evaluations about whether the information is harmful i.e. whether it is misinformation, virus or spam. Devices also have to determine if users will infringe on privacy by excessively accessing private information stored on the device or leaking any information they access. It should be noted that devices are less able to perceive nuance and cannot rely on intuitive, social notions of trust. However, users still leave a digital trail in device networks with trends and patterns. Devices can use this to determine if people are trustworthy in that \textit{they are authorised} and if they \textit{are causing or mean to cause harm}. Figure \ref{fig: thingtopeople} gives an overview of factors that we considered based on these two perspectives along with the above security issues.

\subsubsection{Trust in the Person's Authorisation}
First, devices need to verify if users are authorised. In Section \ref{subsec: cellIPP2P_env} and Figure \ref{fig: digitalworld}, we discussed that owners have usage relationships with their devices, while in vertical industries, personnel interact with devices. Only non-malicious and real users should be able to pass authentication to gain access. However, passwords can be leaked and users can become malicious. Trust management can help devices detect if a user is malicious in nature. If a user has a malicious identity, they should not be given access.

Knowledge factors elaborate on a truster's identity. Using this understanding of the trustee's identity, devices can identify if the features of the user is unusual or harmful. Unusual or harmful behaviour indicates the user is in fact, not authorised to use the device.
\begin{description}
  \item[Trustee authentication] A first obvious step to determine trustworthiness is to require authentication. If a user cannot pass authentication, they are not authorised to access the device because their identity is inauthentic or they have malicious intentions.
  \item[Trustee features] Trustee features possessed by device users can also give clues about whether a user is unreliable. These are features tied to the identity of the user. For example, if personnel in a vertical industry network happens to have very low rank within the plant but is requesting a very high level of access, this could suggest the person trustee has gained access to the device for malicious purposes.
\end{description}

Every user has identifying features in the way they use devices and what information they store. Experience factors determine if identifying features follow an unusual pattern.
\begin{description}
  \item[Profile inconsistency] A user typically has identity-related features stored within a device. However, through usage, additional data is collected. This data could be inconsistent with previously stored data. Confirmation would be useful here to determine if the inconsistency was malicious or accidental or well-intentioned. If a users identifying features are inconsistent with before, the same level of authorisation cannot be granted less the user has changed.
\end{description}

Reputation factors describe how people trustees are connected to the rest of the digital world. These connections inform if the trustee has a malicious identity in the digital world based on their connections with other devices in the network.
\begin{description}
  \item[Trustee authority] Trustee authority is important in digital environments like vertical industries where multiple users may access a single device. Trustee's of different authority level have different levels of access within a network. A trustee that attempts to access a level beyond their workplace authority should be flagged so their behaviour can be accounted for with other potentially suspicious behaviour.
  \item[Connections] Trustees are socially connected in the real world. These connections may be used to authorise access to devices. For example, a friend may unlock their personal devices for their friends to use temporarily; Low level personnel may be given access to devices temporarily by other workers. These situations should be accounted for to avoid wrongfully writing users as untrustworthy.
\end{description}

\subsubsection{Trust in the Person's Behaviour}
Devices should trust users based on their behaviour, especially the typicality of behavior and harmfulness of actions. For example, in vertical industries illustrated in Figure \ref{fig: digitalworld}, devices can receive malicious instructions to perform harmful actions or they can receive harmful or incorrect data. Devices that receive malicious instructions could damage infrastructure; misinformation could affect device's computation and output; harmful information like spam jams the network and service-systems. The more harmful a user's actions are, the less likely they are trustworthy. It is possible the current user is not the real device owner, has been given access but has malicious intentions or the device owner is making a mistake. Devices should defer from executing unusual instructions until further confirmation.

Devices can determine if user's behaviour is atypical or harmful using knowledge factors. Knowledge factors aid experience evaluation to enhance trustworthiness evaluation. Understanding the environment and activities undertaken in the environment, give insight into how potential harmful or atypical behavioural patterns of trustees are.
\begin{description}
  \item[Activity risks] When determining if the actions taken by a trustee are trustworthy, it is important to consider the level of concern required. Some actions have bigger impact than others. For example, in smart agriculture, if a very large amount of water is to be released into the plantation, this presents a very large risk of flooding the entire plantation and damaging all the plants. With such extreme and harmful actions, greater caution is required so the device should require additional verification to make sure the action is well-intentioned or defer the action completely.
  \item[Security requirements] In some digital environments, the general level of trustworthiness could be very low. The sample space of users has a larger number of untrustworthy users, so the odds of encountering a untrustworthy user is higher. Higher security requirements should be standard for all trustees to reduce the chances of an untrustworthy person gaining access to the device network.
\end{description}

Trust in a user's behaviour requires evaluation of their activity. Experience factors are necessary here to evaluate based on experience if a user will or is behaving harmfully or unusually. This helps verify their identity and that their identity is non-malicious.
\begin{description}
  \item[Behavioural anomaly] Behavioural anomaly is an important factor to consider. Some devices store activity history and are able to track behavioural patterns of users. If a user begins to exhibit anomalous behaviour, this suggests the device has changed hands. If there is no reason for this change, the authenticity of the user should come under suspicion. For example, if devices record sudden traffic to highly unusual, particularly dangerous sites, this suggests a malicious user has hacked into the device. This is also true of vertical industries where multiple personnel can interact with devices. If a personnel makes an anomalous request that could damage the device network, caution should be taken to ensure personnel are not malicious or making errors. Alteratively, since devices can also record data from people, if data greatly deviates from trend, the trustee is potentially untrustworthy. For example, if a device receives data about traffic congestion from a user that is a significant distance away from the reported location or the reported traffic congestion occurs at an highly unusual timing, the data and trustee are likely unreliable.
  \item[Past performance] Like in other types of trust, performance also plays a role in determining trustworthiness. Performance determines how capable and willing a user is to handle devices properly. For example, if a user tends to download viruses easily, it shows the user is probably incapable of identifying and avoiding bad links or using the device maliciously. If personnel frequently logs incorrect data, them and any future data they log could be flagged as untrustworthy.
\end{description}

When users release information into device networks, information from within the device network can also give information about how harmful and incorrect the information is. This sort of evaluation relies on reputation factors about the trustee's behaviour.
\begin{description}
  \item[Informational conflict] In digital environments where information is collected from multiple users, data about an observation or instructions at any time can come from multiple people. If one person's information is conflicts with that of a large majority of other people, this information and trustee should be flagged. Alternatively, if data is corroborated by a highly reliable person or device, this indicates the information is highly reliable so, the trustee demonstrates good performance and is reliable.
  \item[Activity abnormality] Activity abnormality uses generalised information about user behaviour to determine if a particular user's behaviour is suspicious. For example, if a user downloads software from an unusual site where users typically do not download such software, devices can flag this behaviour as unusual. The device can then look out for additional signs that may suggest the user is untrustworthy and intends to behave in a malicious manner.
\end{description}

\subsection{Thing to Thing Trust}
\begin{figure*}[t]
  \centering
  \includegraphics[width=\textwidth]{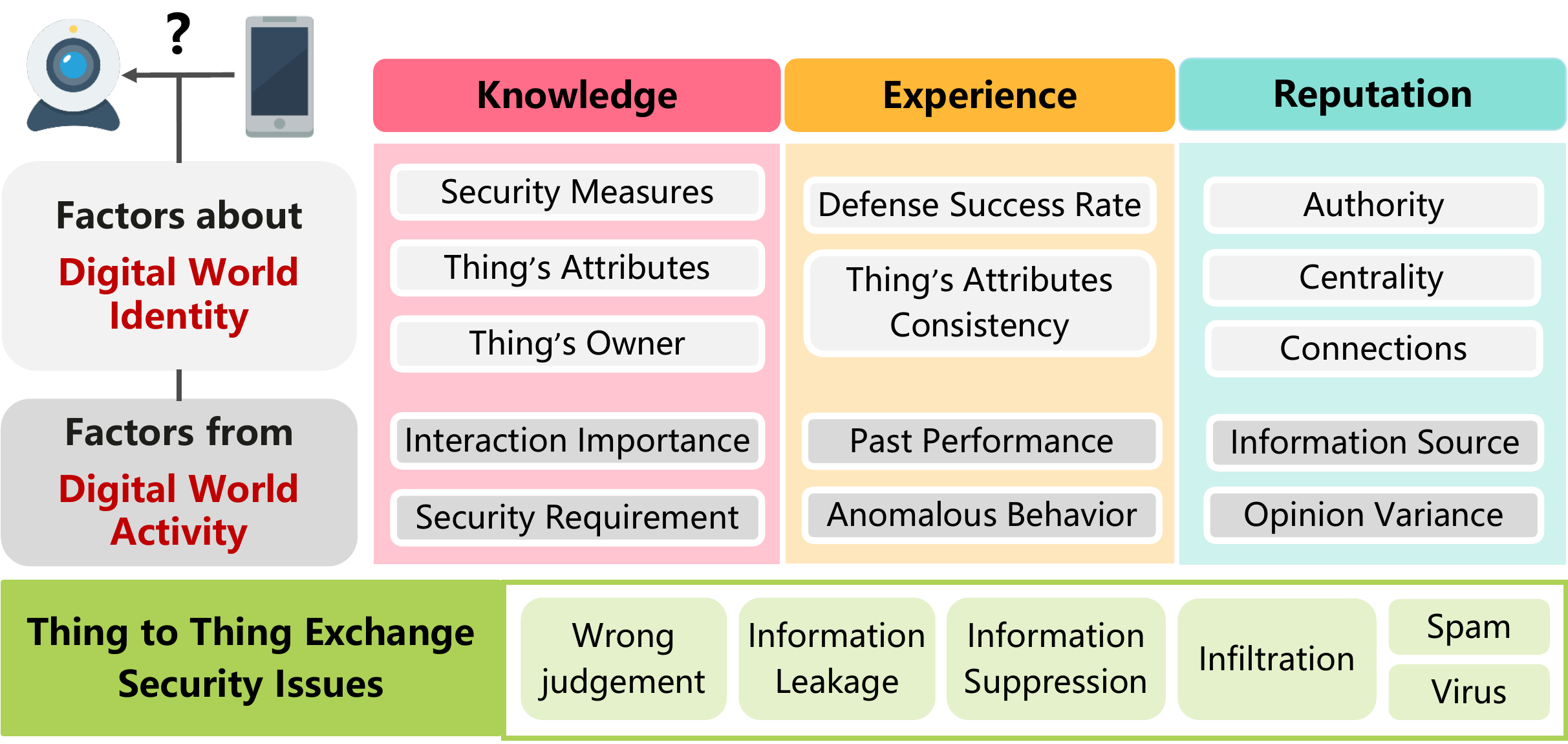}
  \label{fig: thingtothing}
\end{figure*}

In many digital networks, devices pass information on to other devices to form a large network of automated communication. In these networks, devices can transfer poor computational output to other devices, leak personal information to other devices, fail to pass information on to other devices or pass on harmful information such as spam or virus. To do any of these things, malicious devices infiltrate the network. Thing to thing trust is trust between two non-sentient entities. To address security challenges in device networks, thing to thing trust is required. Devices need to believe the trustee thing is \textit{inherently trustworthy} or is \textit{at least trustworthy within the impending interaction}. Factors for each of these instances is outlined in Figure \ref{fig: thingtothing}

\subsubsection{Thing's Inherent Trustworthiness}
Devices can evaluate if trustee things are inherently trustworthy. In Section \ref{subsec: cellIPP2P_env}, Section \ref{subsec: IoT} and Figure \ref{fig: digitalworld}, devices supported by widespread telecommunication infrastructure and open device networks, can receive data from unfamiliar devices. Malicious devices can exploit this openness to infiltrate networks and cause disturbances. For example, they can send a virus-infected file to a node in a wireless sensor network or use communication infrastructure to randomly attack personal devices in peer-to-peer networks. Truster devices thus need to ensure that trustees are inherently trustworthy and do not have intentions to carry out malicious attacks.

To determine if a device's identity is trustworthy, we can use knowledge factors. Knowledge factors give insight into the device and its surrounding social or physical environment. These give insight into the purpose of deploying the device and in turn, its future actions. In addition, knowledge about the digital environment also provides insight into the likely identity of the device so we consider those as well.
\begin{description}
  \item[Security measures] Whether a device is secure against malicious attacks determines its dense capability against malicious attacks and incidents that compromise its performance. What security measures are employed determine this security. For example, a device that has anti-virus software should be able to defend against viruses transmitted from other devices. Therefore, they are less likely to pass on the virus to other devices when transferring files if they have never been compromised. This goes both ways. If a device lacks the appropriates software to defend against malicious attacks, the device should set higher requirements for trustworthiness values to exercise more caution.
  \item[Thing's attributes] A device's characteristics also give information about whether it is capable enough to be considered trustworthy. A device that is older and more primitive might be more prone to mistakes and vulnerable to attacks, particularly if its software is not regularly updated. Therefore, the information from the device is less reliable, making the device less trustworthy.
  \item[Thing's owner] A thing can also be connected to a owner. If two owners know each other, even the two devices have never interacted, a relationship can still be established between the two devices. For example, if two device owners wish to share an Internet connection, the two devices can connect via tethering even if the two devices have never connected before. That the two devices are owned by friends is a characteristic that influences mutual trustworthiness of both devices.
\end{description}

Experience factors provide evidence about the pattern of behaviour, any anomalies and exposure history of trustees. Using this information, trusters can infer the identity of a device --- whether it is capable or willing to cause harm. Inferring this provides information about device's future actions and thus, their trustworthiness.
\begin{description}
  \item[Defense success rate] Despite its best efforts, devices may find that they still encounter malicious attacks. These attacks could compromise the device making it such that the actions of the device affect other devices it interacts with. Then, the success rate of the device and whether or not the device has been compromised in the past becomes relevant before choosing to interact with it. For example, if a device has experienced information leakage before, a device that wishes to pass information through that node may choose not to do so; if the cause of the information leakage is still present, it will leak the truster device's information.
  \item[Thing's attribute consistency] Another experience feature about device identity is whether or not the attributes of the device remain consistent with time. If a device has a particular characteristic, this characteristic needs to be demonstrated with time and its various actions. Otherwise, the device is behaving unusually. For example, a device such as a laptop has relatively high computational power. However, if it often fails at simple processing tasks, this could suggest the laptop has been compromised so, it is behaving oddly.
\end{description}

Finally, we look at reputation features that give insight into the likely intentions of the device. By relying on reputation, devices overall network behaviour acts as additional evidence in case current experience is insufficient to reliable infer about a device's intentions.p
\begin{description}
  \item[Authority] In some device networks, selected devices have greater authority. They may be large professionally-deployed communication infrastructure, government deployed units or officially authorised nodes in the network. Since these devices are verified, the information they distribute is deemed instantly and definitively trustworthy. For example, in vehicular networks, vehicle nodes can directly trust road-side units (RSU) because they are officially deployed to disseminate important information about traffic conditions.
  \item[Centrality] Centrality is important due to the network-nature of device networks. Between two devices, there are common devices that both truster and trustee have interacted with. Depending on the trust values and the proportion the common to total acquittance devices, how information flows between truster and trustee device can be visualised. If there are a significant proportion of common trusted acquaintances, the indirectly flow of reliable information between the devices is significant. So, just like in people to people trust, centrality also matters.
  \item[Connections] Connections influences the formation of trust between devices in two ways. First, since trust is transitive, a truster can use a trusted node to exchange with a trustee, forming an indirect connection for information exchange. Second, if a device has formed many disreputable connections, they are likely to have been compromised on multiple occasions or are also a malicious device. Therefore, it is likely the device is unwilling or incapable of performing well in an exchange, making it untrustworthy.
\end{description}

\subsubsection{Trust in Behaviour Harmfulness}
Devices often do not have sufficient information about the inherent trustworthiness of a device based of its identity. An alternative would be to simple evaluate if a device's incoming action is trustworthy. This is analogous to determining if a trustee will do harm and requires evaluation of the trustee's past and current actions. In Section \ref{subsec: IoT}, it is hard to determine if a device in an outdoor wireless sensor network is inherently untrustworthy. Even if the device is originally trustworthy, at any time the device can topple over and record data wrongly or be physically manipulated by a malicious entity to send inaccurate data. In such environments, no device can be inherently trustworthy. Instead, truster devices can only evaluate if the incoming data and interaction with the device is non-harmful.

To carry out such evaluation, knowledge factors about the interaction and environment are very useful. Devices can infer the likelihood of a harmful interaction and to what extent the interaction would be harmful. This aids the decision to proceed with an interaction.p
\begin{description}
  \item[Interaction importance] Whether an interaction is important significantly affects the degree of caution the truster should employ. If a truster device is exchanging highly sensitive data, the trust threshold should be much higher to ensure that trustee nodes do not either fail to pass on necessary information or pass on misinformation. Data with a large impact on the real world can also be considered important. For example, if a device such as a vehicular node releases information about traffic conditions, this information will direct traffic along different roads. If the information is wrong, intentionally or otherwise, this could cause significant traffic congestion. Therefore, only highly trusted nodes should be allowed to release such high-impact information.
  \item[Security requirement] If there are many untrustworthy nodes in a network, truster devices should also exercise greater caution. This applies like discussed before since most device networks are a type of network in some form or another. Therefore, each node interacts with other nodes in the network. If the network contains a large proportion of malicious nodes, it is more na\"{i}vely probable that a randomly chosen node is malicious. Therefore, extra caution needs to be taken during any interaction.p
\end{description}

To determine if a current interaction will be successful, it is useful to gather past evidence about the trustee device. This requires experience factors which indicate whether the device has been compromised and whether it will behave harmfully.
\begin{description}
  \item[Past performance] Naturally, like in all types of trust, the performance of a trustee is important. A trustee's performance directly reflects their willingness and capability to cooperate in exchanges. If a device frequently provides erroneous information, it is likely the device is either compromised or too old so, it can no longer be trusted. It would thus be better for the entire network to simply ignore the device as its past performance does not bode well for the device's future performance.
  \item[Anomalous behaviour] Since "things" refer to devices, which while not static, cannot interpret nuanced, subjective information, we rely on the trend of behaviour to determine if the behaviour of a device can be considered suspicious. If a device typically behaves in a certain way, any behaviour that fails to follow this trend suggests a change in its surroundings or tampering by some party. If there are no events that may suggest this change, this should arouse suspicious that the device has been compromised.
\end{description}

Finally, to evaluate if actions from a device are harmful, we can rely on the behaviour and information being transmitted. In particular, the reputation of the action within the network is very telling in determining if the incoming action is trustworthy, even if not much can be directly known about the trustee or their behaviour.
\begin{description}
  \item[Information source] Since devices form parts of a network, the information typically travels in paths and cycles. This means that information can be tracked to its original location. If the original source of the information is unreliable, this could suggest that the information is, by proxy, unreliable. Furthermore, the act of propagating this information to the truster suggests the distributing device has poor connections, reducing their trustworthiness.
  \item[Opinion variance] Finally, trusters can receive information from multiple devices in a device networks. Potentially, the device may receive similar information from multiple sources or if a group of devices all manage the same network, the data collected from all devices need to be logically consistent. This data collected as a group acts as a type of corroborating evidence. If the data collected by the trustee device deviates too greatly from what is expected, this indicates that the information and thus, the device is untrustworthy.
\end{description}

\section{Criteria for Digital Trust Modelling} \label{sec: criteria}
\begin{figure*}[t]
  \centering
  \includegraphics[width = \linewidth]{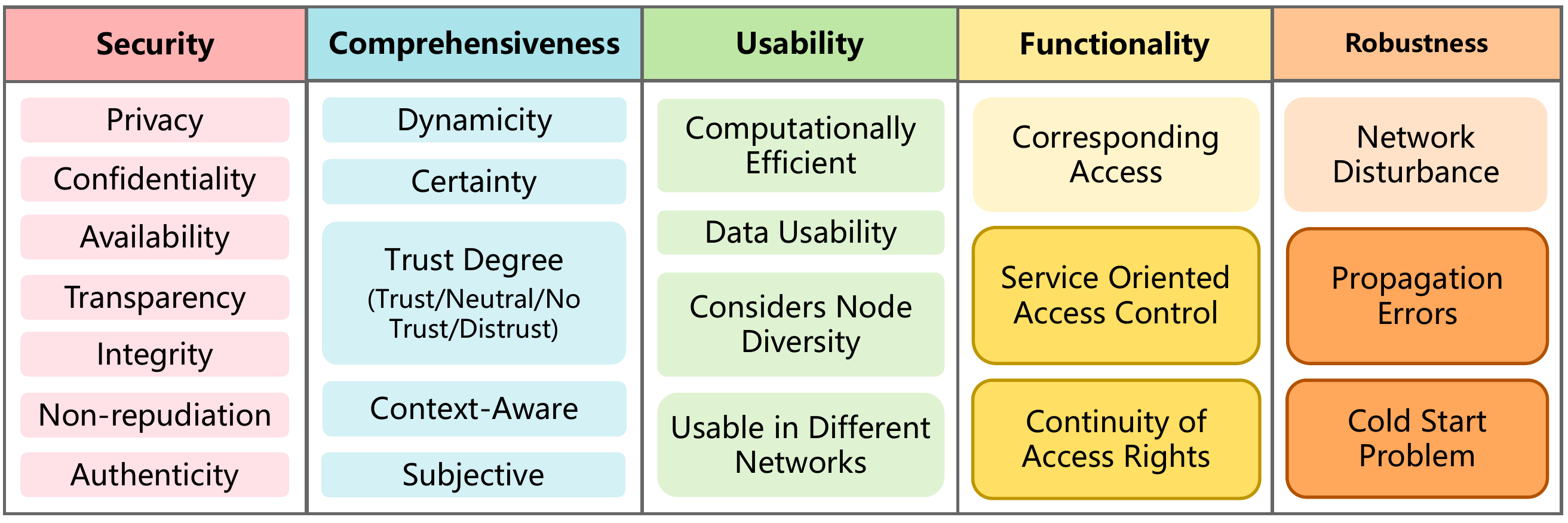}
  \caption{Table of rubrics categorised into general concepts.}
  \label{fig: criteria}
\end{figure*}

In Section \ref{sec: digitalworld}, we laid out the different digital environments and mentioned some trust modelling challenges. In this section, we discuss different rubrics and criteria to address basic trust modelling targets and these additional challenges. They are outlined in Figure \ref{fig: criteria}.

\subsection{Security}
A good trust model is secure meaning it is safe for use and can defends against threats, attacks, errors and back doors. There are many requirements for a trust model to be secure.
\begin{description}
   \item[Privacy] A secure trust model preserves privacy \cite{trustsecurity,mlsurvey,iotsurvey}. This means it allows users to select the type, method and to whom their personal information can be accessed. Trust management schemes must be able to prevent the invasion of privacy by malicious attackers while itself minimising the use of personal information while managing trust. Entities within the network \emph{and} trust models should not able to retrieve user's information without permission of the user. Privacy is necessary to the security of the trust model as lack of privacy could make user feel uncomfortable, resulting in harm, even if the company has no intention of misusing the data. That the user cannot confirm such misuse will not occur is sufficient to cause distress.
   \item[Confidentiality] Confidentiality is a concept highly related to privacy and is too, necessary in trust management \cite{VANETsurvey,iotsurvey}. Confidentiality involves the prevention of excessive collection and leakage of personal data. Like in privacy, this means that both the digital service \emph{and} the trust model should maintain confidentiality. Maintaining confidentiality means trust management should only collect relevant data and cannot leak data to any entities outside those that absolutely need it. Personal data leakages are harmful to users as they could reveal sensitive information, causing real life harmful effects on the user.
   \item[Availability] A trust model is secure if it is available whenever it is needed \cite{VANETsurvey}. This means that attackers are not able to cause disturbances to the network such that users are not able to access its services. Trust is needed here to ensure that when users interact, interactions that attempt to jam and reduce availability, such as spam and virus attacks, are filtered. The process of trust modelling also cannot compromise the performance and capacity of the network.
   \item[Transparency] Trust management should be transparent such that its processes and the information collected are open to each user. The system also needs to be accountable for any failure to protect digital services and errors made during the trust management process. Transparency also means that there should not be any back doors. A lack of transparency in managing trust makes users wary of digital environments.
   \item[Integrity] Integrity is an obvious requirement of any trust management system \cite{iotsurvey,VANETsurvey}. A trust management that has integrity is able to prevent malicious or harmful interactions and behaviour. Trust is needed to determine and measure when an entity is malicious. Then, these entities need to be punished or removed accordingly. Integrity protects the digital environment and its interactions from encountering and being excessively harmed from bad interactions.
   \item[Non-repudiation] Non-repudiation refers to holding all users accountable for their actions \cite{VANETsurvey}. So, when malicious or harmful interactions occur in the network, the relevant perpetrators can be tracked and punished for the interaction. Trust is needed for this as the trust degree should track these harmful behaviours and accurately reflect them in numerical values. Entities with low trust values should not be able to interact in the environment.
   \item[Authenticity] Non-repudiation and its implementation is closely linked to authenticity. Authenticity means that user's identities can be verified. Being able to authenticate users prevents them from being able to take on multiple identities to take the blame for their harmful actions in the network. Authenticity of the network ensures that trust values and the management system is in fact effective. If users were able to create false identities when their trust values fell too low, the trust values would have no meaning. Therefore, authenticity is a necessary requirement for all trust models \cite{VANETsurvey}.

 \end{description}

\subsection{Comprehensiveness}
Comprehensiveness is only partially considered by most trust models. A comprehensive trust model considers all the different dimensions and aspects of trust modelling and is able to adjust to account for these aspects. Comprehensive trust models should fulfill the following requirements.
\begin{description}
  \item[Dynamicity] Dynamic trust models are models that consider the time-varying nature of trust and are necessary as trust is itself, dynamic \cite{onlinesocialsurvey}. Many models are dynamic by considering the age of evidence, recommendations and trust values. Older evidence is less reliable as it is less reflective of the entities future behaviour. However, writing off old evidence can result in a lack of evidence.
  \item[Certainty] Trust, being an inherently complicated concept, carries an inherent uncertainty, as discussed in Section \ref{subsec: digitaltrust}. Models that are comprehensive should consider this uncertainty in its trust values and decisions. This means that when evidence is lacking or contradictory, the metrics used --- be it the trust value itself or a separate certainty indicator --- need to reflect this and factor it into trust decision-making.
  \item[Trust Degree] A comprehensive trust model obviously requires some representation of trust \cite{onlinesocialsurvey}, be it binary, discrete or continuous. However, since trust is complicated this trust degree must also reflect nuanced understandings of trust. Trust models should also be able to account for the differences between "no trust" and "neutral" and reflect them accordingly in the model. Moreover, "no trust", "neutral" and "distrust" are not necessarily the same. "Distrust" implies that the user is repelled by the node whereas "no trust" or "neutral" could simply be that there is insufficient information or conflicting information respectively about the user's trustworthiness.
  \item[Context Aware] Since trust is by nature a highly situational concept, trust models need to consider the different situations and how they affect trust \cite{iotsurvey,mlsurvey}. Not only do they have to consider the different situations, they have to do so appropriately, choosing suitable contextual features that inform a node's trust the most. A trust model that can consider this is holistic and sufficiently comprehensive for implementation in real life.
  \item[Subjective] As mentioned, trust differs greatly from person to person. Truster's requirements need to be holistically considered, factoring in all the relevant truster features to determine the most fitting trust value and decision for the truster. A trust model that can adjust to fit its users needs is sufficiently comprehensive for all users \cite{onlinesocialsurvey, mlsurvey}.
\end{description}

\subsection{Usability} \label{subsec: usability}
Many trust models today claim to accurately detect malicious users and attacks, however, few models consider whether in they are usable and implementable in real-life digital environments. We break down how trust models can in fact become usable for real world digital environments.
\begin{description}
  \item[Computationally Efficient] As mentioned in Sections \ref{subsec: cellIPP2P_env} and \ref{subsec: IoT}, computational efficiency is a huge limiting factor in trust model performance. Therefore, trust management schemes need to carefully manage time and storage complexity as well as the convergence of algorithms \cite{mlsurvey,attackdefense,VANETsurvey}. However, complex and time-consuming algorithms can still be used in appropriate digital environments. Networks containing primitive devices cannot utilise high computational consumption algorithms but trust models for online social networks, that utilise large, central servers and processes might be able to.
  \item[Data Usability] As mentioned in Section \ref{subsubsec: directtrust}, ratings, trust values and other indicators are commonly assumed to exist in digital environments. This is not necessarily true. Furthermore, feature based knowledge modelling in Section \ref{subsubsec: directtrust} may not be able to access the necessary inputs due to unavailability of data or privacy concerns. To even be implemented, trust models need to consider what data is available and if their factors can indeed be computed. Proxies or alternatives need to be found for uncomputable factors.
  \item[Considers Node Diversity] For trust models to be usable they also need to be able to account for individual nodes. This is related to the cold start problem. Trust models that rely on factors such as experiences, face the issue where some nodes lack sufficient experience to have high direct trust values. Such new nodes find themselves unable to interact but unable to perform the necessary exchanges to raise their trust values sufficiently. Trust models should find alternative methods to calculate trust that can account for all the possible different users in the network.
  \item[Usable in Different Networks] Trust models need to be usable in different types of networks. This means that trust management and its calculation methods need to be able to work in networks, even if they contain cycles \cite{frankwalter}, are sparse or have other problematic graph characteristics. Particularly, graph theory related methods and path-finding algorithms may not converge, or collapse whenever cycles or sparse networks are involved. Alternative modelling methods need to be considered accordingly.
\end{description}

\subsection{Functionality}
A trust model must be functional in that it provides an appropriate decision-making framework. Many trust models often aim to output a trust value. While this is important, the trust values should be interpreted into \textbf{corresponding access} for the trustee. In other words, trust models need to provide corresponding access to the different trust values and levels. There are several problems access control and functionality must address.
\begin{description}
  \item[Service-oriented Access Control] Corresponding access control needs to be service oriented \cite{trustsecurity}. This means that the number of thresholds, how narrow the bands and the strictness for different access levels need to be tailored according to the service. Service here should additionally consider provision-context, the user and device  to which the service is being provided.
  \item[Continuity of Access Rights] Access control also needs to identity when a user is not trustworthy enough to continue their access rights \cite{trustsecurity}. Their rights then need to be revoked accordingly. Whether a user is trustworthy enough depends on the application, even if a single malicious act were sufficient to revoke access rights. Therefore, even for trustworthy agents, their trust values should be tracked, constantly evaluating trustees.
\end{description}

\subsection{Robustness}
Robustness has to do with the trust management systems ability to function and provide appropriate trust values despite \textbf{network disturbances} and system-errors. Network disturbances may occur in soft-security distributed digital environments. Therefore, in the transmission of trust related data, trust management may find themselves vulnerable to disturbances where for example, nodes propagate wrong trust values or a malicious attack occurs. Trust management schemes need to be robust enough such that these trust values are still available and the digital environment functioning, no matter these disturbances. Trust management faces two key types of disturbances.
\begin{description}
  \item[Propagation Errors] Errors can also include propagation errors such as natural errors made in the transmission of information between nodes. Trust systems should be able to account for potential transmission errors of trust values and adjust the trust value or certainty accordingly. This ensures digital environment and security availability at all times.
  \item[Cold Start Problem] The cold start problem mentioned in Section \ref{subsec: usability} is relevant here. A trust management system is robust if it can counter instances of insufficient information about the trustworthiness of a node and still produce a good, usable trust value. This can be interpreted as the system being comprehensive enough to be robust even under problematic circumstances.
\end{description}

\section{Attacks on Trust Models} \label{sec: attacks}
\begin{figure*}[t]
  \centering
  \includegraphics[width = \textwidth]{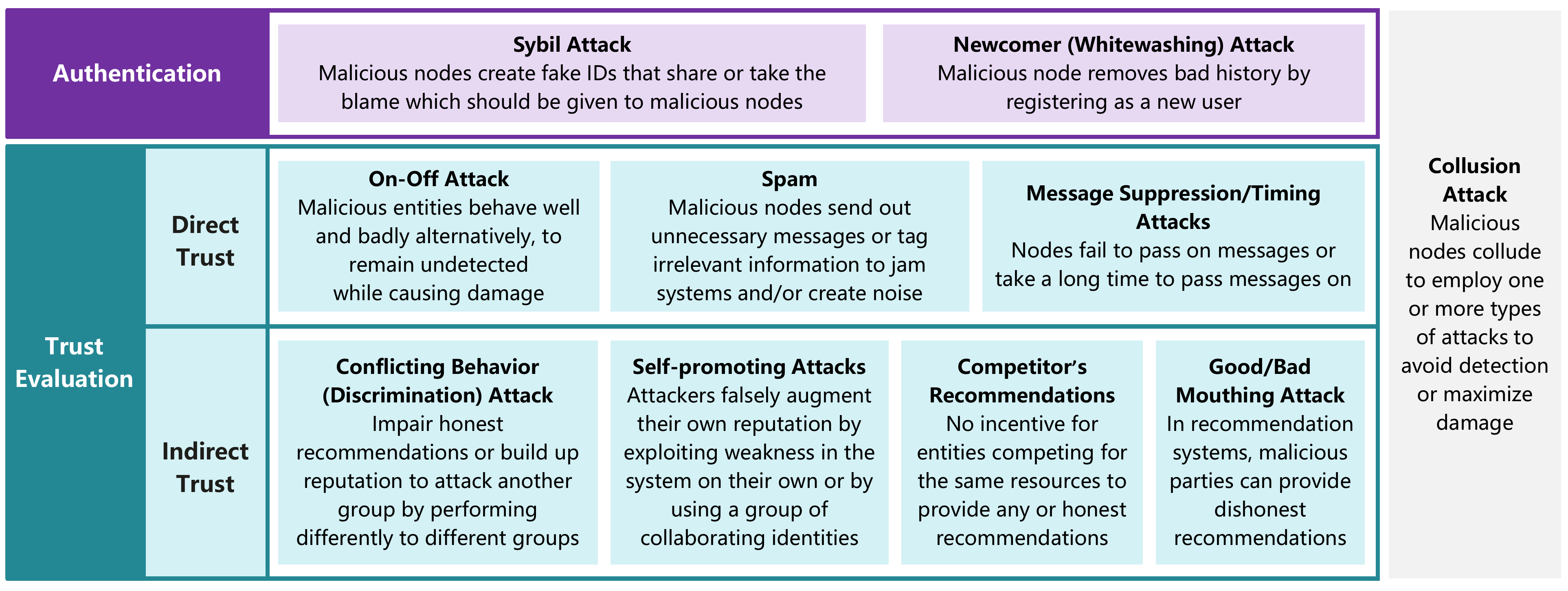}
  \caption{Table of standard trust security threats}
  \label{fig: risks}
\end{figure*}

Security protocols are informed by known attacks on trust management systems. A secure trust management system is able to resist the attacks seen in Figure \ref{fig: risks}. We describe each attack, potential repercussions and how preventing them helps secure trust models.

\begin{description}
  \item[Sybil attack] In Sybil attacks, malicious nodes create fake IDs that share or take the blame which should be given to malicious nodes \cite{attackdefense, trustsecurity, VANETsurvey, onlinesocialsurvey}. Sybil attacks are carried out by malicious nodes by creating many fake identities that then each attack honest nodes. Instead of the single node taking on all the blame for the collection of attacks, the malicious nodes all from the same user share the blame so that their individual trust values drop more slowly.  This way, the collection of malicious nodes are able to launch more attacks before being individually detected. A system that is ensures authenticity of users, non-repudiation and integrity is able to resist Sybil attacks. Users are authenticated so that multiple identities cannot be created without at least being able to trace the all nodes back to the same identity. The authenticated identity is then held responsible for attacks of any attacks by its malicious nodes and removed before other nodes can launch attacks.
  \item[Newcomer attack] In newcomer attacks, malicious node removes bad history by registering as a new user \cite{trustsecurity,attackdefense}. The newcomer attack is slightly different from the Sybil attack in that malicious attackers that have accumulated bad reputations on the system are able to leave the system and return, erasing history of their behaviour so they can launch attacks on honest nodes again. Like the Sybil attack, authentication, non-repudiation and integrity help resist newcomer attacks. Authentication traces the identity of each newly created node to its original creator and are able to carry the bad reputation from the previous node to this new node. The attacker is thus unable to launch attacks despite its new identity.
  \item[On-off attack] In on-off attacks, malicious entities behave well and badly alternatively, to remain undetected while causing damage \cite{trustsecurity}. On-off attacks are attacks launched via direct interactions. Malicious nodes perform well for a while to gain users trust before then attacking the user by performing poorly. They may repeat this again to gain back the users trust before attacking, remaining undetected to maintain trust values above a certain level. A trust model that maintains non-repudiation and integrity will be able to identify malicious nodes performing on-off attacks, punish them and prevent them from attacking.
  \item[Spam] Spam occurs when malicious nodes send out unnecessary messages or tag irrelevant information to jam systems and/or create noise. Spam is another type of attack that occurs in direct interactions. Such spam is typically seen in digital environments with tagging systems \cite{socialspam}. Spam is harmful as it distributes and introduces large amounts of irrelevant, if not harmful, content into networks. Non-repudiation and integrity requirements target this attack so that trust management systems that are secure can identify when a node is distributing large amounts of irrelevant content and remove the node and their content as needed.
  \item[Message Suppression/Timing Attacks] In message suppression attacks, nodes fail to pass on messages and in timing attacks, nodes take a long time to pass messages on \cite{VANETsurvey, attackdefense}. Message suppression and timing attacks are particularly damaging to time sensitive and distribution reliant networks such as vehicular networks or wireless sensor networks. When attackers launch message suppression and timing attacks, they fail to or take a long time to pass messages along. This could result in failures to inform target users about necessary information in time or at all, causing harm to the functioning of the entire network.
  \item[Conflicting Behaviour Attack] In conflicting behaviour attacks, malicious nodes impair honest recommendations or build up reputation to attack another group by performing differently to different groups \cite{trustsecurity}. Conflicting behaviour attacks are damaging to reputation systems. Users destroy the reputations of honest users by behaving positively only with them. By doing so, they build up a good reputation with this select group of users who then recommend the malicious nodes to other honest nodes. This results in a)  the reputation or recommendation nodes being tarnished and b) the truster nodes being harmed from interactions with malicious nodes due to poor recommendations.
  \item[Self-promoting Attack] In self-promoting attacks, attackers falsely augment their own reputation by exploiting weakness in the system on their own or by using a group of collaborating identities \cite{attackdefense}. Self-promoting attacks can occur in poorly designed and poorly authenticated recommendation systems. In systems that are poorly designed, malicious users can either falsely augment their own reputation by recommending themselves or create and collaborate with other users to augment their own reputation. By doing so, malicious users can increase their trust values via indirect trust to perform harmful actions.
  \item[Competitor's Recommendation] In competitive environments, there is no incentive for entities competing for the same resources to provide any or honest recommendations \cite{trustsecurity}. Competing recommendations are particularly problematic in e-commerce platforms where buyers compete for limited products and sellers compete for limited buyers. When there are limited resources, buyers and sellers have no incentive to provide good and honest recommendations \cite{zeinab}. Giving false recommendations results in entities making poor decisions while not giving any recommendations results in users having to act on insufficient information or not acting at all. The lack of incentive to recommend could result in the collapse or stalemate of systems where there are insufficient direct interactions, particularly in new digital environments. Integrity and non-repudiated systems can counter poor recommendations while systems that meet the availability requirement are able to prevent collapse due to lack of recommendations.
  \item[Good/Bad Mouthing Attack] Good/Bad mouthing attacks are attacks where malicious entities provide dishonest recommendations \cite{mlsurvey}. Bad mouthing attacks makes it difficult for trusters to find interactions to serve their needs and trustees to gain the trust of trusters. Good mouthing attacks increase the indirect trust values of malicious nodes and cause honest trusters to interact with malicious entities. Recommendation systems that are flooded with such attacks will drive users away or cause users to no longer rely on recommendations. Systems that meet integrity and non-repudiation are able to resist such attacks.
  \item[Collusion Attack] Lastly, collusion attacks are attacks that use a combination of the above attacks \cite{onlinesocialsurvey}. This allows users to apply maximum damage to the digital environments. A fully secure trust system must also be ensure that the digital environment is not vulnerable to a combination of attacks and that defence against one attack does create vulnerabilities to another.
\end{description}

\section{Modelling Methods} \label{sec: methods}
The criteria discussed can be met by employing the right methods for modelling. In this section we discuss existing methods used in trust modelling, their theories and how they have been used. First, we present an overview of each of the steps in trust modelling. For each of the steps, the existing methods that have been used are illustrated and coloured by the mathematical field they are under.
\begin{figure*}[t]
  \centering
  \includegraphics[width=\linewidth]{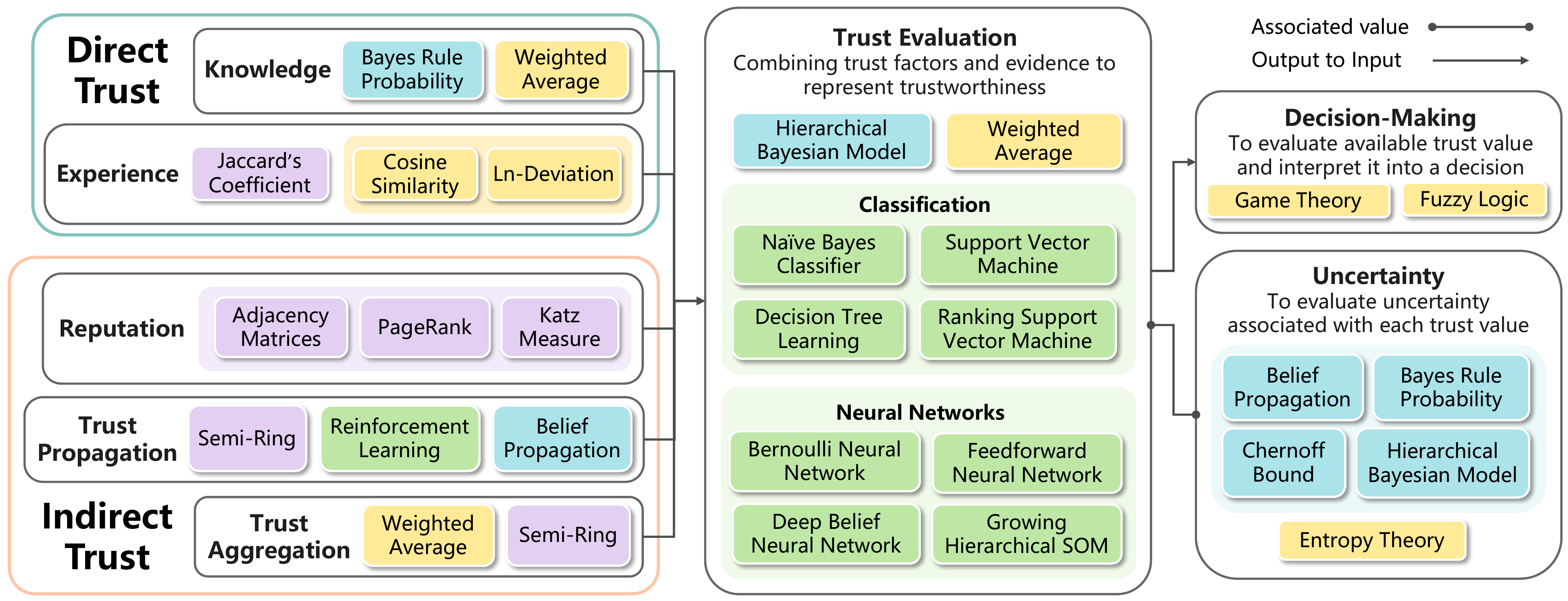}
  \caption{Overview of trust modelling process and methods in each step.}
  \label{fig: methodsmodel}
\end{figure*}

\subsection{Basic Methods}
Simple trust models are models that employ basic mathematical constructions. Heuristics are employed and introduced in these equations so that their trends represent trust-related processes in digital environments. There is no fixed way to construct equations, some are adapted from known and common basic constructions.

\subsubsection{Weighted Average}
The key component of many simple models are usually weighted averages. Weighted averages, like averages, combine values of some set and normalises this sum by the number of elements to achieve a representation of the set that considers all its elements. In weighted averages, however, different values are weighted differently, to provide a more reflective combination. Weighted averages are calculated
\begin{equation} \label{eq: weightedavg}
    \bar{x} = \frac{\sum^n_{i=1} \omega_i x_i}{\sum^n_{i=1} \omega_i}
\end{equation}
where $\omega_i$ are weights for each $i$-th factor, $x_i$. There are $n$ factors being considered. Being able to apply heuristics to weights in trust modelling is advantageous in considering factors that do not have fixed mathematical constructions to represent their meanings. This is useful to trust as the actual magnitudes of ratings tend to be subjective.

\paragraph{Combining Trust Factor}
In \cite{info-trust, zhang08, kangxin, hassanvector, nitti, Che2015, mokhtari, FIREtrung}, weighted averages were used to balance factors to output a trust value. Most trust models do not define weights or leave it to implementers or device owners to decide the weight of each factor \cite{info-trust,zhang08,hassanvector,nitti,Che2015,mokhtari,FIREtrung}. This allows users to personalise their trust model according to their needs and makes them responsible for weighting the factors based on what they know about themselves.

However, some models may instead define weights and/or additionally adjust the weights provided by users. For example, in \cite{kangxin}, dynamic confidence factors were used to weight direct and indirect trust such that with sufficient interactions, greater, if not all, weight would be given to direct trust values. Where $N^{T}_{ij}$ is the number of direct interactions between users $i$ and $j$, the dynamic confidence factor is
\begin{equation}
  \alpha_{ij} = \frac{N^{T}_{ij}}{N^{T}_{ij}+c} \qquad \text{or} \qquad
  \alpha_{ij} = 1 - \beta^{N^{T}_{ij}}
\end{equation}
where $c>0$ and $0<\beta<1$ are parameters that can be adjusted by the user. These user-defined parameters help describe how reliant on direct interactions  the user wishes to be. The two equations are both monotonically increasing with limit at infinity $1$ but have different trends. The dynamic confidence factor can be chosen based on the needs of the digital environment.

In \cite{hassanvector}, confidence and time help adjust the user-defined weights for indirect trust. Confidence is modelled such that the more interactions and recommenders there are, the higher the confidence. This increase in confidence can be modelled with any monotonic increasing function. To account for time, \cite{hassanvector} considered the number of time intervals passed since recommendations were received from each recommender. The longer the time lapsed, the smaller the weight assigned.

\paragraph{Aggregate Experiences}
Weighted averages are also used to aggregate experiences \cite{zhang08, mokhtari, FIREtrung, zeinab}. This helps collect all the evidence (from interactions) and combines them into a single representative value. These values can then be used as part of more complex methods. In \cite{zhang08}, $x_i$ in Eq. \ref{eq: weightedavg} would be the number of positive interactions at each time window, $N_{pos}$. Time-based weighting was carried out by simply using a \emph{forgetting factor}, $0<\lambda<1$, exponentiating using time intervals so that older intervals would have smaller weight. The weighted average representing positive interactions was defined
\begin{equation}
    \bar{N}_{pos} = \sum^N_{i=1} N_{pos} \cdot \lambda^{i-1}
\end{equation}
where $i$ is each time-window. Alternatively, if time intervals are not appropriate, time passed $t_i$ after each interaction $N_i$ can also be used to as a weight for each $i$-th interaction \cite{FIREtrung}
\begin{equation}
  \omega_i = e^{-\lambda t_i}
\end{equation}
so that as time passes, the weight given to older experiences will decrease exponentially. Contextual features has also been used as weights. The weights measure how similar the $i$-th experience is similar to the current experience and would take on $\omega_i$ in equation \ref{eq: weightedavg}.

\paragraph{Aggregate Trust Values}
Weighted averages are also used to aggregate trust values \cite{Mohtashemi, REGRET, sun06, dai, nitti}. Trust values can be that describing paths or individual nodes in a network. Typically they are weighted using the trust values of intermediaries so that $\omega_i$ is the trust value about the intermediary(advisor) and the factor is the advisor's opinion about some other intermediary or the trustee.

\subsubsection{Relational Measurement} \label{subsubsec: relationalmeasures}
Relational measures include similarity measures, deviation, correlation measures and other simple constructions. and are simple constructions meant to represent how far two variables, are related, whether positively or negatively. This relation can give insight into whether these two users are likely to trust each other.

\paragraph{Similarity Measures}
Similarity measures are useful to measure how similar two users are in terms of their opinions. It is generally believed that the more similar two users are the more likely they are to trust each other. Cosine similarity was used in \cite{DSNN, ZOLFAGHAR2011833} to measure how similar two ratings were. When ratings are involved, such as in review systems on e-commerce platforms, the similarity between the ratings of two individuals about the same objects can reflect the level of trust in each other. For example, if two users rated the same set of $n$ products and the ratings were collected into vectors $\textbf{a}$ and $\textbf{b}$ for each user respectively, the cosine similarity would be
\begin{equation}
  \cos(\theta) = \frac{\mathbf{a} \cdot \mathbf{b}}{\|\textbf{a} \| \cdot \|\mathbf{b} \|}
\end{equation}
This way the actual magnitude of the scores are recorded and only the relative differences between the way the two users rate will be considered. For example, a user could have rated two products with the vector $(2, 5)$ while someone else rated those same products $(4, 10)$. The reviews may have very different magnitudes but their opinion towards the products are similar (that the first one is worse than the second). Cosine similarity captures this similarity in rating patterns rather than the total magnitudes of the ratings.

\paragraph{Deviation Measure} \label{par: deviation}
$\ln$-deviation is also used to measure how unusual some variable for a particular node is \cite{info-trust,chen2019}. $\ln$-deviation is adapted from mean-log deviation used in income-inequality which is similar to standard deviation except that $\ln$ is applied to reduce the range of very wide-ranging features. For some variable $X(i)$ for user $i$, $\ln$-deviation is given as
\begin{equation}
  lndev(i) = -\ln \left( \frac{X(i) + 1}{\max_{j \in N}\left( X(j) \right) + 1} \right)
\end{equation}
where $N$ is the total number of nodes in the network. The degree to which the variable for the node deviates from some standard (in this case the maximum) could be an indicator of the impact of their actions \cite{info-trust} or the abnormality of responses or behaviour in the network \cite{info-trust,chen2019} and thus, an indicator of trustworthiness.

Deviation was also used in Zeinab trustworthiness reputation scheme \cite{zeinab}. The trustworthiness of an advisor is measured by the competency and willingness of the advisor to provide good recommendations. Competency of the advisor was measured using a standard deviation method. The reliability of ratings was computed using the evidence-based certainty equation introduced by Wang in \cite{certaintymethod}. Then the standard deviation of the credibility of ratings for different sellers was computed to represent the uncertainty in the seller's ratings. This helps account for the reliability of the advisor despite sellers that discriminate in their behaviour by aggregating the general rating behaviour of the advisor for common sellers.

This deviation was multiplied by the dishonesty of the advisor. This dishonesty was also computed using standard deviation and was represented by the successful and unsuccessful interactions with the seller that the advisor has had. Beta probability described in Section \ref{subsubsec: basicbayesian} was used to measure while accounting for uncertainty the expected rating of the seller. The output is then formulated as the expected rating by the advisor or buyer about the seller. The standard deviation was used on this expectation for common sellers to reflect whether or not ratings provided by the advisor align with those experienced by the buyer. If the ratings do not align, the standard deviation would be high reflecting a high level of dishonesty.

Basic difference between ratings was also used and modified using heuristics to compute the willingness of an advisor to provide good ratings. The difference in the rating by the advisor and all other ratings over time was iteratively computed over different time intervals. This was coupled with integration over the exponentiated form (to give greater weight to larger rating differences). The final integration reflects the rating behaviour of the advisor relative to the buyer. A large accumulated difference in weighting reflects that the advisor is less willing to provide accurate recommendations. This difference was also weighted by the ratio of good recommendations provided by the buyer to the recommender previously that were truthful. This captures the intuition that the advisor will react in kind to how the advisor was treated by the buyer in the past.

\subsubsection{Fuzzy Logic} \label{subsubsec: fuzzylogic}
Unlike Boolean logic where "true" and "false" are the only values allowed, fuzzy logic allows multiple values to be assigned to better represent situations where a statement cannot be determined absolutely true or absolutely false. These intermediate logical values take the form of fuzzy sets such as "not very true", "somewhat false" and so on. Translating observable real life quantities to numeric values can be done via the membership function. These membership functions output the degree to which the quantity is belongs to some defined fuzzy set. Fuzzy logic operators and rules act on values attached to fuzzy sets. These rules help map quantities to a fuzzy set output. The fuzzy output can be translated back into a numerical value if necessary \cite{fuzzytheory}.

The entire process described has been implemented by Song for evaluating trust in grid computing based on numerical type features \cite{Song04fuzzytrust}. In their trust management model, numerical values of features defense capability and job success rate were fuzzified. Using membership functions, the membership degree of the self-defense capability feature in the fuzzy sets "low", "medium" and "high" was determined; the membership degree of the job success rate in the fuzzy sets "very low", "low", "medium", "high", "very high" was determined. Based on intuition, fuzzy rules and logical operators were defined and used to map the fuzzy set values to fuzzy set values describing trust. The possible trust fuzzy sets were "very low", "low", "medium", "high" and "very high", and the fuzzy rules and operators help obtain membership degrees for trust value in each of these fuzzy sets. Using the membership degrees and the inverse membership function, a numerical trust value can be obtained by aggregation.

The concept of membership functions was implemented in \cite{decisiontree, fuzzynn}. The decision tree, explained in Section \ref{par: decisiontree}, is trained by iterating through the different splitting conditions, using a greedy algorithm, to determine the best condition that minimizes the error between derived trust values and actual trust values. These splitting conditions can be used to classify users into trustworthy and untrustworthy classes. However, as highlighted in the motivations for fuzzy logic, continuous attributes are rarely completely in one (intermediate) category or another. Fang addressed this by using membership functions to partially allocate users into each category for each continuous attribute. By iterating through different thresholds in the training process the decision tree can be trained to be more flexible by determining the best threshold to partially allocate users in different categories.

A similar method was employed by Wu to train a fuzzy neural network \cite{fuzzynn}. By preprocessing continuous inputs, Wu categorised continuous inputs into sets to reflect human interpretation on whether or not a particular value is "low", "just so-so", "high" or "very high". Therefore, the neural network can process instead the categories and the degree to which the user belongs in a particular attribute instead of the raw values themselves. This allows human interpretation within the neural network. The fuzzy operators and rule was replaced with the neural network. This process is described in Section \ref{par: fuzzynn}.

\subsubsection{Game Theory}
The iterated prisoners dilemma problem in game theory has been applied to trust modelling for reputation systems in e-commerce markets by Zeinab \cite{zeinab}. In the Prisoners' Dilemma game, two players have to decide whether to cooperate or defect. If both players cooperate, the reward is $R$; if both players defect, the punishment is $P$; if only of the players defect, the player that defected receives reward $T$ while the other receives punishment $S$. These payoffs satisfy $T > R > P > S$ and $2R > T + S$ \cite{ipdtheory}. In the iterated Prisoners' Dilemma game the two players repeat the prisoner's dilemma multiple times and can remember and react to past actions \cite{ipdtheory}.

In e-commerce markets, buyers must balance between competition due to limited resources and not being able to discover good sellers if they report untruthfully. Zeinab adapted this iterated Prisoners' dilemma game to address this aspect of reputation systems in e-commerce markets \cite{zeinab}. In the reputation system, an advisor can choose to cooperate or defect by providing or not providing good recommendations about sellers. The rewards and payoffs from the Prisoner's dilemma can be defined analogously here. Using the defined payoffs, an expected payoff from continuing the pattern of cooperation and defection over multiple interactions can be defined. This long term payoff computation considers the time remaining in the e-commerce network and the trustworthiness value of the advisor. This trustworthiness value of the advisor is determined by the competency and willingness of the advisor to provide good recommendations \cite{zeinab}. Standard deviations and difference values were used and are described in Section \ref{par: deviation}.

\subsubsection{Entropy Theory}
Entropy theory aims to quantify the uncertainty associated with a random variable. To do this, entropy theory quantifies information gained or uncertainty as $I(p)=-\log p$ because \cite{entropytheory}
\begin{itemize}
  \item $I(p)$ is monotonically decreasing. This reflects that as the probability of an event occurring increases, we are less uncertain of the possible outcome so, uncertainty decreases.
  \item $I(p) \geq 0$ which reflects that uncertainty is always non-negative.
  \item $I(1) = 0$ which reflects that events that almost surely occur do not have any uncertainty.
  \item $I(p_1, p_2) = I(p_1) + I(p_2)$ which reflects that the uncertainty of two independent events is the sum of the uncertainty from either event.
\end{itemize}
Then, the amount of entropy associated $H(X)$ with a random variable $X$ is the expected amount of uncertainty with the variable based on all the outcomes and the probabilities of those outcomes. Therefore, entropy is defined \cite{entropytheory}
\begin{equation}\label{eq: entropydef}
  H(X) = \mathds{E}\left[ I\left( p \right) \right] = - \sum^n_{i=1} P\left(x_i\right) \log P\left(x_i\right)
\end{equation}
where the base of $\log$ depends on the application. In trust, base $2$ is usually used. Then, given the random variable $X$ that an agent cooperates with probability $p$, the entropy of the agent cooperating is useful in representing the degree of uncertainty in the agent. Explicitly, this entropy is
\begin{equation} \label{eq: trustentropy}
  H(p) = -p \log_2 p - (1-p) \log_2(1-p)
\end{equation}
using Equation \ref{eq: entropydef}. Entropy has been used as a trust value \cite{sun06, dai} and as a weight \cite{jayasinghe}.

\paragraph{Entropy as a Trust Value}
Trust is often understood as the probability of an agent behaving cooperatively. However, trust may not increase linearly with probability of cooperation. Using entropy as a trust value, probability-based uncertainty associated with the agent is used to represent trust. The trust value is usually defined \cite{dai,sun06}
\begin{equation}
  T(p) =
  \begin{cases}
  1-H(p) & \text{if $0.5 \leq p \leq 1$} \\
  H(p) - 1 & \text{if $0 \leq p < 0.5$}
  \end{cases}
\end{equation}
so that for more extreme probabilities of cooperation (extremely high or extremely low probability), the rate of change of certainty is very much higher than if the probability is not as extreme. At less extreme probabilities (around $0.5$), the rate of change in certainty is much slower. Notice that the entropy is flipped in this definition to represent certainty instead of uncertainty. Trust using this definition of certainty models a truster who
\begin{itemize}
  \item For extremely low probabilities of cooperation, trust drops rapidly. If an agent already has a low probability of 0.2 of cooperating and they drop further to 0.1, their trust value drops to a more disproportionate extent to show that the truster is disproportionately more certain the more extremely low the probability of cooperation.
  \item For mid-value probabilities, certainty in the agent, and therefore trust, changes slower. So if an agent's probability of cooperation increases from 0.5 to 0.6 (0.4 to 0.3), the probability is still fairly low (not that low) and the difference to the truster between the two values might not be very large. So, the 0.1 increase (decrease) results in a disproportionately smaller increase (decrease) in trust.
  \item For extremely high probabilities of cooperation, trust increases rapidly. If an agent with very good probability of performing well can further improve their performance, this continued improvement reflects very well on them. Therefore, the increase in certainty in the agent increases disproportionate.
\end{itemize}

\paragraph{Entropy as  Weight}
Entropy has also been used as a weight in trust modelling \cite{jayasinghe}. Using entropy from Equation \ref{eq: trustentropy} as a weight, more weight is given to interactions that have less extreme probability values. For each interaction, there is a probability of the interaction outcome being positive. If by entropy, the interaction carries more information, more weight should be given to the interaction when aggregating all experiences.

Jayasinghe did this in his model for IoT devices \cite{jayasinghe}. Adapting his method, suppose a truster has had $c_1 , \dots, c_n$ interactions with a trustee, for each interaction $c_m$ there is a probability $p_m$ of the interaction succeeding. Using Equation \ref{eq: trustentropy}, the uncertainty-weighted performance of the trustee denoted $CFD$ is
\begin{equation}
  CFD = \sum^n_{m=1} \frac{c_m}{t_m} H(p_m)
\end{equation}
where $\frac{c_m}{t_m}$ is the fraction of total time spent by the trustee for each interaction and helps to further weight the interaction based on its duration. In $CFD$, if an interaction had roughly $50/50$ chance of succeeding, its success was more uncertain and so any information gained from the interaction would be greater, by simplifying some of the uncertainty. Therefore, more weight is given to the interaction outcome and duration.

\subsection{Graph Methods}
In graph methods, we discuss the available methods to capture the structure od a network, globally and locally for each node. These graph methods are particularly useful because many digital networks can be simplified into graphs with nodes and edges. The existing relationships and transfer of any information offers large amounts of information that can be modelled into features for trust. Otherwise, graph methods are very useful for inferring trust values about other nodes based on the trust values of surrounding nodes or along paths.

\subsubsection{Network Features}
First, in graph methods for trust modelling, of key interest is how to describe a node's structural position in a network and relationships to other nodes. We call these network features and discuss how graph methods can and have been used for modelling of such network structural features.

\paragraph{Network Overlap}
The degree to which two individuals in a network have common "friends" is an indicator of how significant a role the other plays in each other's network. The larger the role, the more significant the role and so the more likely there will be trust between the two users. There are two possible representations in trust modelling: centrality and Jaccard's coefficient.
Centrality is measured by a simple mathematical concept in a number of models \cite{nitti,jayasinghe}. Centrality of trustor $i$ and trustee $j$ is calculated as
\begin{equation}
    c_{ij} = \frac{\left| K_{ij} \right|}{ \left| N_j \right|}
\end{equation}
where $K_{ij}$ is the set of common friends and $N_j$ is the set of friends that$j$ has. Then, $c_{ij}$ represents the degree to which $i$ and $j$ are indirectly connected. $N_j$ is necessary for normalization. For the same $\left| K_{ij} \right|$, a larger $N_j$ means that the number of common friends is proportionally smaller. Thus the overlap in connections is not as significant.

Jaccard's coefficient is a similar measure but normalized with the total number of friends \cite{DSNN}. Where $I(i)$ is the set of $i$'s friends, Jaccard's coefficient for trust between $i$ and $j$ is usually
\begin{equation}
  J_{ij} = \frac{\left| I(i) \cap I(j) \right|}{\left| I(i) \cup I(j) \right|}
\end{equation}
Then, Jaccard's coefficient represents the proportion of the friend group considering both users. This means that if either user has a large group of friends, the union set would be large and so the number of shared friends would not be proportionally large enough to warrant a large amount of trust between the two, based on common friends.

\paragraph{Global Reputation}
A more global relational measure, adapted from PageRank, measures the importance of a node in relation to the entire network based on its links. PageRank is an algorithm typically used to rank webpages on search engines. In trust modelling, PageRank is modified so that the rank given to the page is analogous to the authority or popularity of a particular node in the network \cite{chen2019,ZOLFAGHAR2011833,info-trust,DSNN}. This reputation score is given by
\begin{equation}
    R(i) = \frac{1-d}{n} + d \sum_{j \in M(i)} \frac{R(j)}{L(j)}
\end{equation}
where $n$ is the total number of nodes, $M(i)$ is the number of people who trust $i$ and $L(j)$ is the number of nodes that $j$ trusts. $d$ is the damping factor, representing in the case of trust, when the user continues trusting others within the network, instead not trusting anyone. PageRank algorithm for trust outputs a ranking of nodes in the network that considers number and trustworthiness of the truster nodes for any particular trustee node. It is assumed that the trustworthiness of a trustee node is indicated by the number of nodes that trust the trustee.

\paragraph{Adjacency Matrices}
Adjacency matrices are square matrices that represent finite graphs. The $ij$-th entry in the matrix indicates a relationship between node $i$ and node $j$ in the graph. In a weighted graph, like those typically used in trust, the weights will be used directly as entries in the adjacency matrix. Otherwise, the adjacency matrix will be binary with a $1$ entry indicating the presence of a relationship and a $0$ representing no relationship. For trust, the adjacency matrix is most useful for illustrating, in matrix form, the total path weights (for weighted graphs) or total number of paths (for non-weighted graphs) from any one node to another. This representation allows an entry-by-entry visualisation of the type of relationship each node has with every other node. This representation can be further expanded into other metrics.

\paragraph{Katz Measure}
Katz measure is useful in measuring the degree of influence of a node in a network based on all the paths going to the node \cite{Katz1953}. Katz measure was used in \cite{ZOLFAGHAR2011833} as an input feature to a neural network and in \cite{frankwalter} for the indirect trust value. Katz measure of user $i$ is given
\begin{equation} \label{eq: katzmeasure}
    X_{katz}(i) = \sum^{\infty}_{k=1} \sum^n_{j=1} \alpha^k(A^k)_{ji}
\end{equation}
where $A$ is the adjacency matrix of the underlying graph. Then, $(A^k)_{ji}$ counts all the paths of length $k$ from user $j$ to $i$ and does so for all users $j=1$ to $j=n$. This is repeated for paths of all lengths from $k=1$ to $k=\infty$. The "attenuation factor" $0< \leq \alpha \leq 1$ then weights the paths such that the longer the path, the less weight assigned to it. This reflects how trust generally diffuses the more intermediate recommenders there are along a path.
Computing Equation \ref{eq: katzmeasure} is equivalent to finding the column sums for the following matrix \cite{Katz1953, frankwalter}
\begin{equation}
    \bm{X}_{katz} = (\bm{I} - \alpha \bm{A})^{-1} - \bm{I}
\end{equation}
where $\bm{I}$ is the identity matrix. When $\alpha$ is less than the reciprocal of the largest characteristic root of $\bm{A}$, computations can be simplified significantly so that there is no need to compute powers of matrices \cite{Katz1953}. Otherwise, an iterative method proposed by \cite{frankwalter} also offers a more efficient method of computation.

\subsubsection{Semi-rings}
Semi-rings are algebraic structures defined by a tuple $\left( A, \oplus, \otimes, \circled{0}, \circled{1} \right)$ such that for all elements $a,b,c$ in the non-empty set $A$ and for the elements $\circled{0}, \circled{1} \in A$, the following conditions hold true \cite{george,STAR}
\begin{itemize}
  \item $\oplus$ is commutative, associative and $\circled{0}$ is the additive neutral element:
    \begin{gather}
        a \oplus b = b \oplus a \\
        (a \oplus b) \oplus c = a \oplus (b \oplus c) \\
        a \oplus \circled{0} = a
    \end{gather}
  \item $\otimes$ is associative, $\circled{1}$ is the neutral element and $\circled{0}$ is the absorbing element:
    \begin{gather}
        (a \otimes b) \otimes c = a \otimes (b \otimes c) \\
        a \otimes \circled{1} = \circled{1} \otimes a = a \\
        a \otimes \circled{0} = \circled{0} \otimes a = \circled{0}
    \end{gather}
  \item $\otimes$ distributes over $\oplus$:
    \begin{gather}
        (a \oplus b) \otimes c = (a \otimes c) \oplus (a \otimes b) \\
        a \times (b \oplus c) = (a \otimes b) \oplus (a \otimes c)
    \end{gather}
\end{itemize}
The properties of the semi-ring can be interpreted to correspond to human interpretations of trust. These interpretations are described in \cite{george} and \cite{hypergraph} and demonstrate the usefulness of semi-rings for trust modelling.

\paragraph{Semi-rings for graphs}
Semi-rings have been proposed by \cite{george} for trust-modelling between nodes in graphs.

Typically, the set $A$ in semi-rings used for trust modelling are Cartesian planes of trust and confidence values. Each point in the Cartesian plane is denoted $(t, c)$ for trust and confidence respectively. The semi-ring operator $\oplus$ serves to combine trust values of nodes along a path and while $\otimes$ serves to aggregate the trust values across different paths.

In \cite{george}, two semi-ring constructions were proposed. In both semi-ring definitions, the additional properties were imposed to reflect properties of trust. These properties are
\begin{itemize}
  \item $a \otimes b \leq a, b$ to reflect that opinions propagated along a path are limited by the trust values of nodes along the graph and
  \item $a \oplus b \geq a, b$ to reflect that opinions aggregated across paths are more information rich so they should be greater than individual opinion values.
\end{itemize}

The first construction is the path semi-ring.  In this semi-ring, trust and confidence values each come from the domain $[0,1]$. Therefore, the trust semi-ring would be $S = \left( [0,1], [0,1], \oplus, \otimes \right)$ and the operators are defined \cite{george}
\begin{align}
  \otimes: & \quad \left( t_{ik}, c_{kj} \right) \otimes \left( t_{kj}, c_{kj} \right)= \left( t_{ik} t_{kj}, c_{ik} c_{kj} \right) \label{eq: georgepathplus} \\
  \oplus: & \quad \left( t^{p1}_{ij}, c^{p1}_{ij} \right) \oplus \left( t^{p2}_{ij}, c^{p2}_{ij} \right) \\
  &=
  \begin{cases}
    \left( t^{p1}_{ij}, c^{p1}_{ij} \right), & \text{if } c^{p1}_{ij} > c^{p2}_{ij} \\
    \left( t^{p2}_{ij}, c^{p2}_{ij} \right), & \text{if } c^{p2}_{ij} > c^{p1}_{ij} \\
    \left( \max (t^{p1}_{ij}, t^{p1}_{ij}), c^{p1}_{ij}\right), & \text{if } c^{p2}_{ij} = c^{p1}_{ij}
  \end{cases} \label{eq: georgepathtimes}
\end{align}
where $t_{ij}$ indicates the trust from $i$ to $j$ and $c_{ij}$ indicates the confidence in this trust value. The superscript $p_kp$ indicates that the corresponding value refers to that for a path $k$. In this construction, only the trust value of one path is used at the end to decide the trustworthiness of the user.

\cite{george} also proposed an alternative distance semi-ring construction. The domain of trust values is $\left[ 0, \infty \right]$ and confidence values if $\left[ 0,1 \right]$. Therefore, the semi-ring is $S = \left( \left[ 0, \infty \right], \left[ 0,1 \right], \oplus, \otimes \right)$. The operators were defined
\begin{align}
  \otimes: & \quad \left( t_{ik}, c_{kj} \right) \otimes \left( t_{kj}, c_{kj} \right) = \left( \frac{1}{\frac{1}{t_{ik}}+\frac{1}{t_{kj}}}, c_{ik}c_{kj} \right) \label{eq: georgedistance}  \\
  \oplus: & \quad \left( t^{p1}_{ij}, c^{p1}_{ij} \right) \oplus \left( t^{p2}_{ij}, c^{p2}_{ij} \right) = \left( \frac{c^{p1}_{ij}+c^{p2}_{ij}}{\frac{c^{p1}_{ij}}{t^{p1}_{ij}} + \frac{c^{p2}_{ij}}{t^{p2}_{ij}}}, c^{p1}_{ij}+{c^{p1}_{ij}} \right)
\end{align}
so that the trust and confidence values are considered for multiple trust paths. Both semi-rings behave differently as demonstrated in \cite{george} and should be chosen based on the digital environment.

When considering more than two nodes along the path, the above semi-ring operators proposed by \cite{george} can be extended to iterative computation so that the trust value between the source $s$ and target $t$ is given
\begin{equation}
  \left( t^{p_k}_{st}, c^{p_k}_{st} \right) = \bigotimes_{e_{ij} \in p_k} (t_{ij}, c_{ij})
\end{equation}
where $e_{ij}$ is an edge carrying the trust and confidence values from $i$ to $j$ and $i$ and $j$ are intermediate nodes that from $s$ to $t$ in the path $p_k$.

Similarly, with more than two paths, the final path aggregated trust value between source $s$ and target $t$ is
\begin{equation}
  (t_{st}, c_{st}) = \bigoplus_{p_k \in \text{paths}} \left( t^{pk}_{ij}, c^{pk}_{ij} \right)
\end{equation}
where $p_k$ is the $k$-th path in the set of possible paths from $s$ to $t$.

In \cite{STAR}, it is noted that confidence values are not immediately available to implement the semi-rings proposed by \cite{george}. To obtain certainty values, \cite{STAR} proposes two $c_{ik}$ functions, that takes in the edge $e_{ik}$ between two directly connected nodes $i$ to $k$. The first function reduces the confidence value as paths get longer by defining $c_{ik}$ to be
\begin{equation}
  c_{ik}(e_{ik}) = \alpha, \quad 0 < \alpha < 1
\end{equation}
using a constant $\alpha$ as a decaying factor. Then clearly, by Equation \ref{eq: georgepathplus} and \ref{eq: georgedistance}, the confidence between two indirectly connected nodes decreases the more edges there are between two the nodes \cite{STAR}. This reflects the intuitive understanding that the more intermediate recommenders are required, the more likely information is diluted by different opinions and so the less reliable the propagated information.

Alternatively, \cite{STAR} also proposes a distance-based confidence function defined linearly by
\begin{equation}
  c_{ik}(e_{ik}) = \min(\beta + \gamma d_i, 1)
\end{equation}
or exponentially by
\begin{equation}
  c_{ik}(e_{ik}) = 1-\eta^{d_i}
\end{equation}
where $d_i$ is the degree of node $i$ in the edge from $i$ to $k$ and $\beta, \gamma, \eta$ are tunable parameters. Note that $0<\eta<1$. The underlying intuition is that the degree of certainty of the trust value given by $i$ about $k$ is determined by how reliable $i$ is and the more edges $i$ has, the more reliable $i$ is. Therefore, the greater the degree of $i$, the greater the confidence in the trust value provided by $i$ about $k$.

\paragraph{Semi-rings for hyper-graphs}
Hyper-graphs differ from graphs in that each edge can point from one node to multiple nodes. Hyper-graphs are useful for representing relationships in digital environments where trust can be towards a group of individuals rather than a single entity.

In the trust hyper-graphs proposed by \cite{hypergraph}, trust relationships can be established between one trustor and a group of trustees. The direct edges typically found in a normal graph can be grouped to form an "AND connector". The trust-confidence tuple attached to the edges can be combined to form a single tuple describing the relationship the trustor has with multiple trustees. To form an indirect trust relationship between a trustor and a target group of trustees, the trustor, through intermediate groups, to the target groups. This is called an "AND tree". To compute the trust and confidence values attached to the "AND tree", the semi-ring from Equation \ref{eq: georgepathtimes} in \cite{george} was proposed.

\subsection{Bayesian Methods}
Interactions in digital environments depend on many observable and unobservable variables that play into people's decision making and so, the outcome of any interaction can be perceived as a random variable. Trusting another agent prior to an interaction can then be measured by the believed probability that some interaction in a trust environment would result in a positive outcome. This interpretation of trust allows the use of probability theory and inference for trust modelling and decision making. In particular, Bayesian probability is frequently applied \cite{Mohtashemi,TEACYHABIT,dimitri,zhang08,BLADE,kbt,ganeriwal,Che2015,dai,sun06}. Just as humans combine their knowledge and observations to make decisions, Bayesian probability which is founded on Bayes rule combines data and a priori knowledge to produce evidence-based probabilities.

\subsubsection{Bayes Rule Probability Model} \label{subsubsec: basicbayesian}
The most basic Bayesian inference model that will be described in this section has been widely applied in trust \cite{Mohtashemi,zhang08,ganeriwal,Che2015,dai,sun06}. Bayes rule is generally given by
\begin{equation}\label{eq: bayesrule}
  p \left( \theta | \text{data} \right) \propto p \left( \text{data} | \theta \right) p \left( \theta \right)
\end{equation}
which tells us that the distribution of some parameter or random variable given some observations can be given by some combination of a likelihood ($p \left( \text{data} | \theta \right)$) and prior knowledge ($ p \left( \theta \right)$).

Let trustworthiness be the random variable $T$ and there have already been $s$ successful interactions and $n$ negative interactions with the target agent. Bayesian probability can determine the probability of trust given the collection of past interactions as evidence. This is the posterior. In the context of trust, $T$ will be the belief in the trustworthiness of an agent and $(s, f)$ will constitute evidence of successful and failed interactions with the agent, respectively. By Bayes rule, \cite{ganeriwal,Mohtashemi}
\begin{equation}
    P(T|s, f) = \frac{P(s, f|T)P(T)}{\text{Normalization}}.
\end{equation}
The prior and likelihood functions are not fixed. In trust, \cite{Mohtashemi,zhang08,ganeriwal,Che2015,dai,sun06} have all used binomial distributions for the likelihood function and Beta distributions for the prior. Then, the posterior can be computed
\begin{align} \label{eq: simplebayesiantrust}
    P(T|s, f) & = \frac{Binomial(s+f, T) \cdot Beta(\alpha, \beta)}{\text{Normalization}} \\
    & = Beta(s + \alpha, f + \beta)
\end{align}
The binomial distribution, $Binomial(s+f,T)$ is the probability mass function of the number of successes in a series of $s+f$ independent interactions. Each interaction has probability $T$ of success, since the agent's trustworthiness reflects the likelihood of successful interactions. $Beta(\alpha, \beta)$ represents the prior knowledge about trustworthiness of the trustee which can range from $0$ to $1$. $\alpha$ and $\beta$ are the beta distribution parameters and determine the shape of initial distribution of the likely value of $T$. For trust, $\alpha$ can be interpreted to be the prior expectation of the number of successful interactions and $\beta$ will be the prior expectation of the number of unsuccessful interactions.

Suppose there is evidence of $(s,f)$ past interactions and suppose not enough is known to form a prior expectation so we set $Beta(1,1) = Uniform(0,1)$. Then the distribution for the trustworthiness value of the target would be, by Equation \ref{eq: simplebayesiantrust}, $Beta(s+1, f+1)$ \cite{ganeriwal}. Then, when there are additional $s'$ and $f'$ successful and unsuccessful interactions, we can let the previous posterior $Beta(s+1, f+1)$  be the new prior and the likelihood function be $Binomial(s'+f', T)$. Then, the final posterior would be $Beta(s'+s+1, f'+f+1)$. For convenience, we denote total positive interactions to be $N_{pos}=s'+s$ and negative interactions $N_{neg} = f + f'$. The expected trustworthiness value is then the statistical expectation of $Beta(N_{pos}+1, N_{neg}+1)$ which is
\begin{equation}
  \mathds{E}\left[ T| N_{pos}, N_{neg} \right] = \frac{N_{pos} + 1}{N_{pos} + N_{neg} + 2}
\end{equation}
which is a single consolidated trustworthiness value.

\subsubsection{Chernoff-Hoeffding Bound} In its simplest form, Chernoff-Hoeffding Bound provides probability bounds for the sum of independent random variables. It states that for $X_1, X_2, \dots, X_m$ independent random variables where $0 \leq X_i \leq 1$ for $i=1, \dots, n$, then for $0 < \varepsilon < 1- \mu$,
\begin{equation} \label{eq: chernoff}
  P \left[ \bar{X} - \mu \geq \varepsilon \right] \leq e^{-2m\varepsilon^2}
\end{equation}
where $\mu = \mathds{E} [\bar{X}]$ \cite{Hoeffdingcboundtheory}. Note that $\mu$ is equivalent to the population mean. By definition of variables, we can interpret this inequality as an upper bound for the minimum difference between the sample mean and population mean.

Mohtashemi, Zhang and Zeinab all applied this bound directly to their respective trust models \cite{Mohtashemi,zhang08,zeinab}. In trust, Chernoff-Hoeffding bound is typically applied to determine the necessary number encounters (the value of $m$) to achieve the desired level of confidence. Mohtashemi, Zhang and Zeinab all defined ratings for interactions between a trustor and a trustee to be binary. In other words, each random variable $X_i$ can have a value of $1$ meaning the $i$-th interaction was successful or $0$ meaning the $i$-th interaction was unsuccessful.

By definition, we can see that $X_1, X_2, \dots, X_m$ is in fact a series of $m$ Bernoulli random variables. Let trust be the probability of success of each interaction (i.e. of each Bernoulli random variable) and denote this $\theta$. Since $X_i$ is a Bernoulli random variable, the population mean $\mathds{E} = \theta$ is the expected number of succesful interactions in the long run. In Mohtashemi, Zhang and Zeinab, $\hat{\theta}$, equivalent to the sample mean of interactions, acts as the estimator of trust, $\theta$, between the trustor and the trustee. Applying Bound \ref{eq: chernoff} tells us that for the $m$ past interactions,
\begin{equation} \label{eq: trustbound}
  P \left[ |\theta - \hat{\theta}| \geq \varepsilon \right] \leq 2e^{-2m\varepsilon^2} \leq \delta
\end{equation}
where $\varepsilon > 0$ is a trustee set constant representing the maximum tolerable error between the actual trust value and estimator. The trustee should also define the maximum level uncertainty allowed for error value, denoted $\delta$. For example, a trustee that is only willing to accept a trust value error of $0.05$ and must be $95$ percent certain that the error is within acceptable range will set $\varepsilon = 0.05$ and $\delta = 1-0.95 = 0.5$. We can then manipulate the second inequality in Bound \ref{eq: trustbound} to be
\begin{equation}
  m \geq - \frac{1}{2 \varepsilon^2} \ln \left( \frac{\delta}{2} \right)
\end{equation}
which tells us the minimum number of interactions to achieve the desired level of trust value accuracy with the desired level of confidence. Using the example above, we would need at least $277$ interactions before the desired level of accuracy and confidence is achieved.

\subsubsection{Hierarchical Bayesian Models}
In hierarchical Bayesian models, the relationship between a random parameter and its observations is extended to multiple layers, where each random parameter forms a layer which can have theoretically infinite layers of random parameters above it (hyperparameters). Each parameter layer influences the parameter below it in the same way $\theta$ influences $\text{data}$ in the posterior of Equation \ref{eq: bayesrule}. Bayes rule given by Equation \ref{eq: bayesrule} is extended to multiple layers of random variables to compute the posteriors - the distribution of all random parameters given the observed data in the lowest level \cite{Bayesianmodeltheory}.

Suppose the random variable $\theta$ is of interest and there are multiple instances of this random variable, denoted $\theta_j$ for the $j$-th random variable. Suppose also that each $\theta_j$ has produced a finite vector of observations $\bm{y_j}=\left( \dots, y_{jk}, \dots \right)$. Let the distribution in which the data is observed be $Q$. Then, it can be said that $y_{jk} \sim Q(\theta_j)$. Further, suppose that each $\theta_g$ in fact rises from a common distribution $W$ described by the parameter $\gamma$. Then $\theta_j \sim W(\gamma)$. This process of defining distribution parameters by other random variables can be repeated as many layers as is needed. In this case, suppose the distribution of $\gamma$ is roughly known (e.g. uniform distribution with known start and end distributions), then by Bayes rule, the posterior for all unknown parameters is
\begin{equation}\label{eq: conditionalhierarchy}
  p \left( \gamma, \theta | \bm{y} \right) \propto p \left( \bm{y} | \theta, \gamma \right) p \left( \theta | \gamma \right) p(\gamma)
\end{equation} \label{eq: marginalgamma}
which gives us the distributions of the parameters of interest. It is more likely that the overarching random parameter that determines the distributions of all the parameters is of interest \cite{Bayesianmodeltheory}. Therefore, the marginal distribution of $\gamma$ can be computed
\begin{equation}
  p \left( \gamma, \theta | \bm{y} \right) \propto \int p \left( \bm{y} | \theta, \gamma \right) p \left( \theta | \gamma \right) p(\gamma) d \theta
\end{equation}
when $\theta$ is a continuous random variable. The discrete case is analogously defined. Computations are typically performed stochastically using methods such as Markov Chain Monte Carlo (MCMC) methods \cite{Bayesianmodeltheory}.

This hierarchical Bayesian model can be immediately and simply applied. For example, $theta_j$ describes the random variable where agent $j$ is cooperative and $\bm{y}_j$ is the past experiences that have been had with agent $j$. In this case, $\gamma$ would determine agent behaviour in general, being the parameter in the distribution of agent behaviour $theta$. The posterior of random variable $\gamma|\bm{y}$, computed using Equation \ref{eq: marginalgamma} would thus tell us the distribution of the behaviour of agents given past experiences. One issue with Bayesian models is that the distribution itself is not defined, even if its parameters can be encoded as random variables. Nevertheless, hierarchical Bayesian models have been used for trust modelling.

A more complicated hierarchical Bayesian model, BLADE, was applied by Regan for e-commerce and reputation systems \cite{BLADE}. In BLADE, ratings by advisors and buyers about different sellers are random variables influenced by random variable representing the features of the respective sellers. Since the distribution parameters of ratings and features are unknown, their distributions are determined further by random variables, denoted $\theta$ by Regan, that determine the respective distributions of each feature or rating random variable. Dirichlet distributions were used to describe each of the individual random variables. Time is incorporated into this model by dropping the parameters of the Dirichlet distribution at every time step by a constant factor.

HABIT, a hierarchical Bayesian model proposed by Teacy, takes a different approach with the hierarchy of parameters and chosen distributions \cite{TEACYHABIT}. In HABIT, two different but related models - the confidence and reputation models - are defined. The confidence model describes the likely outcome of a particular interaction between a specific trustor and trustee. The reputation model describes the trustee's global reputation. In the confidence model, the outcome of an interaction between trustor $i$ and trustee $j$ is a random variable, $O_{ij}$. To determine the distribution of the outcome random variable, a random distribution parameter for each outcome random variable, denoted $\theta_{ij}$ , is defined. This random variable is determined by a random vector variable, $\theta_{\cdot j}$ which collects all opinions about trustee $j$. This random vector parameter has an additional random variable that determines its distribution parameters, denoted $\phi$. These variables $\theta_{\cdot j}$ and $\phi$ form the reputation model.

For the confidence model, a vector describing probability of the likely value of the outcome of the interaction was used as the likelihood function. A Dirichlet distribution was assigned as the conjugate prior distribution. Hyperparameters for each Dirichlet distribution determined the shape of the distribution $\theta_{ij}$ and were updated whenever with each direct interaction. For the reputation model, a non-parametric Dirichlet process (not Dirichlet distribution) model. The prior for $\phi$ was fully described by a constant and the set of trustees $\{ \theta_{\cdot j} \}_{j=1}^n$. The posterior for this prior was defined along similar terms. Bayes rule could be used and there would be a final closed form solution. Alternatively, a Gaussian reputation model could be implemented which requires the use of MCMC methods to achieve a closed form solution. In the Gaussian reputation model, $\phi$ is the vector containing all means and covariances which represents general trustee behaviour and how informative reputations sources are \cite{TEACYHABIT}. The likelihood function is then defined using the standard Gaussian probability density function. The prior is then selected to be the normal-inverse-Wishart distribution.

\subsubsection{Dempster-Shaffer Theory} \label{subsubsec: dstheory}
Dempster-Shaffer Theory (DS Theory) is a generalisation of Bayesian probability that maintains the conditioning on observed data and summarising state of belief but removing the need for a global probability distribution assignment \cite{Dempsterorg}. In DS Theory, probabilities are assigned to sets of events rather than each individual mutually exclusive event \cite{gordondsbook,dsevicombi}. This means that evidence collected can be associated with a set of events and assumptions need not be made about the single events within this evidential set \cite{dsevicombi}. For a set of mutually exclusive and exhaustive single events, DS Theory is interested in its power set (set of all subsets) and assigns a mass $m \in [0,1]$ to each element in the power set, via the \textit{basic probability assignment} function (bpa) \cite{gordondsbook,dsevicombi}. It should be noted that the mass for the null set should be $0$ and all the masses should sum to $1$ \cite{gordondsbook}. Formally,
\begin{gather}\label{eq: dsmass}
  m: \mathcal{P}(X) \rightarrow [0,1], \\
  m(\emptyset)=0, \\
  \sum_{A \in \mathcal{P}(X)}=1
\end{gather}
where $X$ is the universal set of events and $\mathcal{P}(X)$ is the power set.

There are 3 functions, besides bpa, that are of interest in trust. They are the belief in event of interest $A \in \mathcal{P}(X)$, the disbelief in $A$ and the uncertainty in $A$. The belief, disbelief and uncertainty functions takes as input the basic probability assignment and are denoted $b(A), d(A)$ and $u(A)$ respectively. The belief in set $A$ represents the total belief about $A$ when all evidence bearing on $A$ has been pooled \cite{gordondsbook}. Disbelief represents the total belief that $A$ does not occur and uncertainty represents the level of uncertainty in the occurrence of event $A$ or $\neg A$ \cite{dsbeliefunbeliefuncertainty}. The following definitions then follow
\begin{gather} \label{eq: dsbeliefs}
  b(A) = \sum_{B | B \subseteq A} m(B), \\
  d(A) = \sum_{B|B \cap A = \emptyset} m(B), \\
  u(A) = \sum_{B \big| \substack{ B \cap A \neq \emptyset \\ B \nsubseteq A }} m(B).
\end{gather}
and demonstrate the method DS Theory takes to generalise Bayesian probability in the assignment of beliefs based on evidence.

Another way in which DS Theory generalises Bayesian probability is in its combination of multiple sources of evidence. Combination rules in DS Theory aggregates these multiple sources of evidence to provide a single meaningful value summarising belief in the events of interest. These multiple sources of evidence provide different belief assessments for the events in the universal set and DS Theory assumes that these sources are independent \cite{dsevicombi}.

Dempster's rule of combination combines evidence using the aggregation of basic assignment values, $m_1$ and $m_2$, in the following way
\begin{align}
  & m_{12}(A) = \\
  & \begin{cases}
  \frac{1}{1-K} \sum_{B \cap C = A} m_1(B) \cdot m_2(C), \quad & \text{ when } A  \neq \emptyset \\
  \qquad m_{12}(\emptyset) = 0, \quad & \text{ when } A  = \emptyset \\
  \end{cases}
\end{align}
where $K = \sum_{B \cap C = \emptyset} m_1(B) m_2(C)$ \cite{dsevicombi}. $K$ represents the basic probability mass associated with conflict. $1-K$ is a normalization factor that serves to completely ignore conflict and attributing the associated probability mass to the null set \cite{dsevicombi}. Combination rules have been shown to provide counterintuitive results \cite{dszadeh}. For this reason, some other types of combinations rules have been proposed \cite{dsevicombi}.

In trust, DS Theory has been used in conjunction with a number of other methods. DS Theory was used by Wang as a precursor to a neural network \cite{DSNN}. In the DS Theory portion of Wang's neural network, a set of representative features were selected from the set of inducing factors. This set of representative features was used as evidence. Then basic belief assignment was carried out using degree to which each input factor belongs to trust and distrust classes. The masses for each class and evidence was combined using a mass combining unit that applied Dempster's rule of combination. The combined masses were then used as input to the fusing units that formed the neural network. How the fusing layers of the neural network work is discussed in Section \ref{par: dsnn}.

An alternative application of DS Theory combined with Beta distributions, as discussed in Section \ref{subsubsec: basicbayesian}, was applied by Zhang and Zhou to their binary trust reputation systems \cite{zhang08,DBNcontextaware}. The possible atomic outcomes defined by Zhang and Zhou were that the trustee was either trustworthy, event denoted $T$, or untrustworthy, event denoted $\neg T$. This means that the power set of interest would simply be $\mathcal{P}(\{T, \neg T\})=\{ \{T\}, \{\neg T\}, \{T, \neg T\}, \emptyset \}$. From general DS Theory, the aim now would then be to assign belief functions to the subset of interest, $\{ T \}$.

With the definitions of belief, disbelief and uncertainty, and because manner in which the set of outcomes is so simply defined, the belief, disbelief and uncertainty functions could be simply given by each of their first equalities in Equations \ref{eq: dstrust}. Then, to assign the mass functions which are necessary to compute belief, disbelief and uncertainty, bpa values have to be assigned. This was done using the number of successful interactions, denoted $r$, and unsuccessful interactions, denoted $s$, as evidence. Borrowing from Beta distributions discussed in Section \ref{subsubsec: basicbayesian}, the expectation of successful and unsuccessful interactions were adapted to define the mass functions. Therefore, the belief, disbelief and uncertainty
\begin{gather} \label{eq: dstrust}
  b(T) = m\left(\{T\} \right) = \frac{r}{r+s+2}, \\
  d(T) = m\left(\{\neg T\} \right) = \frac{s}{r+s+2}, \\
  u(T) = m\left(\{ T, \neg T \}\right) = \frac{2}{r+s+2}
\end{gather}
which were then used by Zhang as a computational tool for reputation scores \cite{zhang08} and by Zhou as a decision making condition in their trust process \cite{DBNcontextaware}.

\subsubsection{Belief Propagation} \label{subsubsec: beliefprop}
Belief Propagation (BP) algorithms coupled with k-Nearest Neighbours Graph (k-NNG), discussed in Section \ref{par: knng}, has been used in trust modelling of online content distribution \cite{Gisel}. Trust modelling using belief propagation requires labels for at least a small number of known entities in the network so that information can be propagated throughout some graph.

Belief propagation is a likelihood updating algorithm that uses Bayes Theorem to propagate the effect of new evidence throughout a directed, acyclic graph called a belief network \cite{bptheory, bptheorycomprehensive}. Using the directed graph obtained from the k-NNG algorithm, the relationship between nodes \circled{A} $\rightarrow$ \circled{B} will be that the trustworthiness of $A$ implies the trustworthiness of $B$. Their edge, is quantified by the conditional probability $P(B_j | A_i)$, where $A_i$ and $B_j$ are the states of variables $A$ and $B$ \cite{bptheory,bptheorycomprehensive}. In the context of trustworthiness, $A, B \in \{0,1\}$ are the trustworthiness variables defined by the indicator random variable in Equation \ref{eq: indicatorvariable}. The edge weight can be interpreted in this context as the probability that $B_j$ is in trustworthy or untrustworthy given that $A_i$ is of trustworthy or untrustworthy.

For a single connected network - there is at most $1$ undirected path between any two nodes - \cite{bptheorycomprehensive} describes an BP algorithm that updates probability values at for each node (variable) in one pass and produces probabilities that are consistent with the axioms of probability theory. Following the general fragment of a singly connected graph in \cite{bptheorycomprehensive}, suppose \circled{A} has parents \circled{B} and \circled{C}. Since the graph is acyclic, the graph above \circled{A} can be partitioned into the subgraph of nodes connected (directly or indirectly) to \circled{B} and the subgraph of nodes connected to \circled{A}. Suppose also that \circled{A} has children \circled{X} and \circled{Y}. Like with parent nodes of \circled{A}, the subgraph that is a child to \circled{A} can also be partitioned into a subgraph of nodes connected to \circled{X} and a subgraph of nodes connected to \circled{Y}.

The data contained in the subgraph of \circled{B} and \circled{C} is denoted $D_{BA}^+$ and $D_{CA}^+$ respectively and the data contained in the subgraph of \circled{X} and \circled{Y} is denoted $D_{AX}^-$ and $D_{AY}^-$ respectively. Node \circled{A} separates \circled{B}, \circled{C} and each of its connected subgraphs from \circled{X} and \circled{Y} and each of its connected subgraphs. Therefore, the effect of $D_{BA}^+$ and $D_{CA}^+$ and $A_i$ trustworthiness data on the children subgraphs is summarised by $A_i$ and we can write \cite{bptheorycomprehensive, bptheory}
\begin{multline} \label{eq: bpsummary}
    P(D^-_{AX}, D^-_{AY} | A_i, D^+_{BA}, D^+_{CA}) \\ = P(D^-_{AX} | A_i)P(D^-_{AY} | A_i).
\end{multline}
The goal of belief propagation in our context would be to find out the probability that $A$ is trustworthy or untrustworthy given all the available data. We can then denote belief for node and state $A_i$ as $\text{BEL}(A_i)$ and define this belief using conditional probability to be
\begin{equation}
    \text{BEL}(A_i) \triangleq P(A_i | D_{AX}^-, D_{AY}^-, D_{BA}^+, D_{CA}^+).
\end{equation}
Following the derivation in \cite{bptheorycomprehensive}, obtain
\begin{multline} \label{eq: bpbelief}
    \text{BEL}(A_i) = \alpha P(D_{AX}^- | A_i) P(D_{AY}^- | A_i) \\
    \cdot \left[ \sum_{jk} P(A_i | B_j, C_k)P(B_j|D^+_{BA})P(C_k|D^+_{CA})\right]
\end{multline}
where $\alpha$ is a normalization constant. Equation \ref{eq: bpbelief} demonstrates how the belief of the trustworthiness of $A$ is determined by causal data from the parent subgraphs, diagnostic data from children subgraphs and the fixed conditional probability matrix that determines how $A$ is affected by its immediate causes: $B$ and $C$.

It can be seen that the children \circled{X} and \circled{Y} then need to propagate $P(D_{AX}^- | A_i)$ and $P(D_{AY}^- | A_i)$ information to \circled{A} and parent nodes \circled{B} and \circled{C} need to propagate $P(B_j|D^+_{BA})$ and $P(C_k|D^+_{CA})$ to \circled{A}. In \cite{bptheory,bptheorycomprehensive} the information propagated from parent nodes is denoted
\begin{equation}
    \lambda_X(A_i) = P(D_{AX}^- | A_i), \quad \lambda_Y(A_i) = P(D_{AY}^- | A_i)
\end{equation}
and that propagated from children nodes is denoted
\begin{equation}
    \pi_A(B_j) = P(B_j|D^+_{BA}), \quad \pi_A(C_k) = P(C_k|D^+_{CA})
\end{equation}
for ease of defining the updating equations.

When some node, for example \circled{A}, receives new information, it needs to update its parent nodes and children nodes. As derived in \cite{bptheorycomprehensive}, the updating function for propagation by \circled{A} to each of its respective parents is
\begin{equation}
    \lambda_A(B_i) = \lambda \sum_j \left[ \pi_A(C_j)\sum_k \lambda_X(A_k) \lambda(A_k) P(A_k|B_i, C_j) \right]
\end{equation}
and the updating function for propagation to each of its children is
\begin{multline}
    \pi_X(A_i) = \alpha \lambda_Y(A_j) \\
    \left[ \sum_jk P(A_i | B_i, C_k) \pi_A(B_j) \pi_A(C_k) \right].
\end{multline}
Belief propagation is touted by its creators to mimic the manner in which people make decisions \cite{bptheorycomprehensive}. It is also guaranteed to achieve equilibrium in time proportional to the network diameter \cite{bptheorycomprehensive} compared to other machine learning methods that may not reach convergence. The calculations necessary for each node is also simple and so is hardware implementable \cite{bptheorycomprehensive} and can address the issues raised in Section \ref{sec: digitalworld}.

It should however be noted that the above calculations only work for singly connected graphs. This is highly unlikely in the context of trust. Between any two people there can be multiple trust relationships. Methods to extend the method for singly connected graphs to multiply connected graphs have been proposed briefly in \cite{bptheorycomprehensive}. Relatively faster variants of the belief propagation algorithm have also been proposed such the Fast Belief Propagation algorithm proposed in \cite{Gisel}.

In Fast Belief Propagation \cite{fabptheory}, the final belief of nodes is approximated by solving for $\bm{b}_h$ in the linear system
\begin{equation}
\left[ \mathbf{I} + a \mathbf{D} - c' \mathbf{A} \right] \mathbf{b}_h = \bm{\phi}_h
\end{equation}
where $\bm{\phi}_h$ is a vector containing the prior beliefs about nodes, $\bm{D}$ is a diagonal matrix, $\bm{A}$ is the adjacency matrix and $\bm{I}$ is the identity matrix. $a = \frac{4h_h^2}{1-4h_h^2}$ and $c' = \frac{2h_h}{1-4h_h^2}$ are constants defined to account for $h_h$ which is the degree of similarity between two nodes. $h_h$ is chosen using conditions in \cite{fabptheory} before solving to ensure convergence.

\subsection{Machine Learning Methods}
Machine learning models are useful for learning different trust factors and assigning weights to them based on data. Machine learning methods vary, each with their unique benefits. The main benefit of machine learning is the construction of trust models based off on data rather than human understandings of trust which may be non-representative and too complex to model.

\subsubsection{Classification}
Classification in trust modelling is generally used to gather multiple factors and use them to classify nodes as trustworthy or untrustworthy. There are a number of different methods each with their individual benefits.

\paragraph{Support Vector Machine (SVM)} \label{par: svm}
SVM is a classification method used in trust modelling \cite{liu08,jayasinghe}. Data was obtained from online communities in \cite{liu08} and an IoT network from a convention from \cite{jayasinghe}. Decisions to interact with reviews on online communities were used as indicators or trust decision on \cite{liu08}. In \cite{jayasinghe}, the trust values were created using k-means clustering so that while there was no explicit trust value in the data set, the effectiveness of the selected features could still be tested to an extent.

In the trust model by \cite{liu08} variables available in the data set were used directly. However, in \cite{jayasinghe} features that were thought to described trust, such as past experience and centrality, were constructed using simple mathematical constructions. In both cases, the selected or modelled features were mapped onto a feature space and SVM applied.

In SVM, the goal of the algorithm is to find a hyperplane that correctly divides data points into trustworthy (represented as $1$) and untrustworthy (represented as $-1$) with maximum margin between the two classes. Where the trust features are represented in the vector $\textbf{x}$ and the trust decision is $y \in \{-1, 1\}$, the set of $n$ trust data points will be
\begin{equation}\label{eq: trustdatapoints}
    (\textbf{x}_1, y_1), (\mathbf{x}_2, y_2), ..., (\mathbf{x}_n, y_n).
\end{equation}
To prevent data points from falling into the margins of the hyperplane and being incorrectly classified, the hyperplane is restricted (for the linear form) such that \cite{svmtheory}
\begin{equation}
    y_i \left( \mathbf{w}^T \mathbf{x}_i -b \right) \geq 1, \quad \text{for all } 1 \leq i \leq n.
\end{equation}
To maximise the split between the data points, the distance between the hyperplanes, given by $\frac{2}{\| \textbf{w} \|}$, should be maximised and so, $\|\textbf{w} \|$ needs to be minimized. Finally, in linear form, the optimal hyperplane that divides the data points correctly with maximal margin will be \cite{svmtheory}
\begin{equation}
    \mathbf{\mathbf{w}^T \mathbf{x} - b = 0}.
\end{equation}

By changing the dot product of vectors $\textbf{u}$ and $\textbf{v}$ to a nonlinear kernel $K(\textbf{u}, \textbf{v})$, the SVM algorithm can be modified to best suit the digital environment \cite{svmtheory}. In \cite{liu08}, radial basis function kernel
\begin{equation}
    K(\mathbf{u}, \mathbf{v}) = \exp \left({-\frac{\| \mathbf{u} - \textbf{v} \|^2}{\sigma^2}} \right)
\end{equation}
was used \cite{svmtheory}. Other knowledge about the digital environment and the way trust decisions are made can also be implemented within the kernel function to make for a more accurate trust model \cite{svmtheory}.

\paragraph{Ranking SVM (RSVM)}
RSVM is a variant of SVM that was used by \cite{RSVM} for establishing trustworthiness of different users on social networks. This variant of SVM was used so that users could be accurately ranked according to features and this ranking used to establish corresponding continuous trust values. The trust values can then be said to be consistent with the rankings established by the features in the trust model.

RSVM for trust modelling can be performed as follows. Suppose, for some trustor $j$ in a data set, there are $n$ trustees to be ranked where each of the trustees have known ranks $r_j^*$. Then, RSVM should aim to find the ranking function $f_{\mathbf{w}}(j)$ such that the resulting rank $r_{f_{\mathbf{w}}(j)}$ and $r^*$ have as few contradicting ordered pairs as possible (discordant pairs) \cite{rsvmtheory}. $\mathbf{w}$ is learned by RSVM and determines $f_{\mathbf{w}}(j)$ which then determines the rank of trustees by the projection of data points in the feature space onto $\mathbf{w}$ \cite{rsvmtheory}. The RSVM optimization problem aims to maximize the number of pairs such that \cite{rsvmtheory}
\begin{equation}
    \forall j \left( \forall (d_x, d_y) \in r_j^* : (d_x, d_y) \in f_{\mathbf{w}}(j) \right).
\end{equation}
This is NP-hard. The solution is approximated by introducing non-negative, slack variables $\xi_{i,j,k}$ and SVM margin maximisation \cite{rsvmtheory} to instead find $\mathbf{w}$ that minimizes  \cite{RSVM}
\begin{equation}
    V(\mathbf{w}, \mathbf{\xi}) = \frac{1}{2} \| \mathbf{w} \|^2 + C \sum \xi_{i,j,k} \\
\end{equation}
subjected to
\begin{equation}
    \forall(j > k): \mathbf{w}^T \mathbf{x}_{ij} - \mathbf{w}^T \mathbf{x}_{ik} \geq 1 - \xi_{i,j,k}, \quad \xi_{j,k} \geq 0
\end{equation}
where $\mathbf{x}_{ij}$ is the normalized feature vector between users $i$ and $j$.

\paragraph{Decision Tree Learning} \label{par: decisiontree}
Decision tree learning has been used in \cite{vehicle20,decisiontree} to classify users into trustworthy and untrustworthy classes. Given a data set with known trust values, the decision tree training algorithm finds the splitting condition that divides the data subset the best for the particular level, at the particular node. This recursive splitting stops when no other splitting condition can add value to the prediction.

Decision trees are useful in considering factors regardless of whether they are binary, discrete or continuous. They are also useful because they allow variables to split in a hierarchical manner, considering certain variables only after some other variable has been considered. The end product is a series of splitting conditions and an order in which to perform this splitting condition that will likely determine if a trustee is trustworthy with the highest accuracy.

\paragraph{Na\"{i}ve Bayes Classifier}
Na\"{i}ve Bayes classifier has been used to classify trustees based on their features using Bayesian probability \cite{chen2019,liu08,homabullying,yuan2006}. Suppose there are features, $x_1, x_2, \dots, x_k$ to be used to classify a trustee, Bayes rule is used to compute the probability that the trustee is trustworthy given the available feature evidence. This is given by
\begin{align}
    p \left( T | x_1, x_2, \dots , x_k \right) &= \frac{p(T)p(x_1, x_2, \dots , x_k|T)}{p(x_1, x_2, \dots , x_k)} \\
                                            &= \frac{p \left( x_1, x_2, \dots ,p x_k, T \right)}{p(x_1, x_2, \dots , x_k)} \label{eq: conditionaldef}
\end{align}
where \ref{eq: conditionaldef} is achieved due to the definition of conditional probability and $T$ is the indicator random variable for the trustworthiness of the trustee, defined
\begin{equation}\label{eq: indicatorvariable}
    T =
    \begin{cases}
        1, \quad & $if trustworthy$ \\
        0,       & $if untrustworthy.$
    \end{cases}
\end{equation}
The denominator is just a normalization factor so we ignore it. To compute the numerator easily, it is na\"{i}vely assumed that all features are mutually independent, conditioned on $T$. Therefore, the posterior is given
\begin{equation}
    p \left( T | x_1, x_2, \dots , x_k \right) \propto p(T) \cdot \prod ^k_{i=1} p\left( x_i | T \right).
\end{equation}
Finally, the classifier will provide the trust decision, $\hat{T}$,
\begin{equation}
    \hat{T} =
    \begin{cases}
        1, \quad & \text{if } p \left( 1 | \mathbf{x} \right) \geq p \left( 0 | \mathbf{x} \right) \\
        0, \quad & \text{if } p \left( 1 | \mathbf{x} \right) < p \left( 0 | \mathbf{x} \right)
    \end{cases}
\end{equation}
where $\mathbf{x} = (x_1, x_2, \dots, x_k)$ is the vector of features  \cite{yuan2006}.

Since trust modelling is a form of social control and modelling where agents act in a highly interdependent manner, the mutual independence assumption is likely too strong. Nevertheless, the classifier and use of Bayes Rule is still applicable. In fact, the use of Bayes rule is beneficial as it allows for a priori knowledge, evidence and the strength of evidence to be accounted for.

\subsubsection{Clustering}
While clustering may not allow for classification of trustees, clustering is still a useful tool for combining and identifying similar agents and entities based on defined criteria for further analysis.

\paragraph{K-means Clustering}
K-means clustering is a common clustering algorithm used to group similar data by their numerical features \cite{kmeanstheory}. Each cluster of data is represented by its centroid, typically a weighted average of all the points within the cluster. The number of clusters, conventionally denoted $K$, is user-defined. The algorithm works by first selecting $K$ initial centroids. Then, the $K$ clusters are formed by assigning each data point to its closest centroid. The centroid of each cluster is recomputed and the assignment process is repeated until the centroid does not change.

Jayasinghe created numerical features describing interactions and connections between trustee and trustor agents in an IoT setting \cite{jayasinghe}. Data however typically does not come with trust value tagged to each interaction. In order to perform further classification, Jayasinghe first performed k-means clustering on the data to group similar interactions into two to three clusters. Doing so helps to distinguish between better and worse performing interactions with respect to the features. Given that the features reasonable represent trust, the clusters then help indicate which interactions were more likely trustworthy or untrustworthy, despite the lack of explicitly recorded data. If there were 3 clusters, one of them could be used to indicate neutrality about trustworthiness.

Obtaining the clusters created a set of labelled data which was then used in SVM for classification. The use of SVM for classification has been discussed in Section \ref{par: svm}. Briefly, SVM helped to find a boundary of features that could distinguish between clusters. This boundary then offers a standard for features to determine if a future interaction should reasonably belong to one cluster or the other.

\paragraph{K-Nearest Neighbour Graphs} \label{par: knng}
Another way to group users together borrows from k-NN classification to build a graph. This graph can function as a method to perform propagation or other trust evaluation methods. Given a set of data, the nearest neighbour of $v_i \in V$ is a point $v_j \in V$, $j \neq i$, with minimum distance from $v_i$. Generally, Euclidean distance is used and to ensure uniqueness of the nearest neighbour, the maximum index when there are ties is used. The edges can then be defined $e(v_i) = \left< v_i, v_j \right> \in E$ and the nearest-neighbour graph would be the tuple $G = (V,E)$ \cite{knngtheory}. The k-NNG graph is simply the nearest-neighbour graph with $k$ edges instead of just one edge.

k-NNG coupled with Belief Propagation (BP) algorithms, discussed in Section \ref{par: beliefprop}, has been used in trust modelling of online content distribution \cite{Gisel}. The idea behind using k-NNG was primarily to group similar entities into a k-NNG so that known labels could be propagated over the rest of the network. This is then useful to determine the trustworthiness of entities even if direct data about them is not explicitly available.

In \cite{Gisel}, the words in online articles were gathered and passed through tensor decomposition to group similar articles. The articles were then represented as nodes in the graph and which were grouped into a k-NNG. Articles that were sufficiently similar to known fake news articles were labelled as fake news. To generate a k-NNG graph, metrics for distance are needed. Besides what was done in \cite{Gisel}, it is also possible to incorporate trust features, such as those described in Section \ref{sec: trusttypes}, within a distance metric like Euclidean distance. The k-NNG graph is then a directed graph $G = (V, E)$ where for each  agent $v \in V$, there are $k$ edges pointing to the $k$ most similar agents denoted $u_i \in V$. It is then known the agents that are feature-wise most similar to each other. This similarity can then be used to make trust decisions, even if data such as quality of direct interactions, is not available.

\subsubsection{Reinforcement Learning}
Reinforcement learning is a type of machine learning method where an agent interacts with the environment in a discrete series of time steps to achieve some goal \cite{rlintro}. In reinforcement learning, the agent may be in \textit{states}, take \textit{actions} and use the received \textit{rewards} to evaluate the quality of their choices \cite{rlintro}. The agent uses the states, actions and rewards to formulate a \textit{policy} which maps states to actions \cite{rlintro,qlearning}. Using this policy, the \textit{return} is the expected future rewards that the agent should aim to maximise \cite{rlintro}. The policy has \textit{value functions} which assigns to states or state-action pairs the expected return if the agent follows the policy \cite{rlintro,qlearning}.

One of the prevalent methods of performing reinforcement learning is Q-learning where the expected value of each action in different states is stored and incrementally updated \cite{qlearning}. The policy is formed from executing the action with the highest expected value \cite{qlearning}. The value function used by Q-learning is a function of the immediate reward and the expected reward based on the new state \cite{qlearning}. This is expressed with
\begin{equation}\label{eq: qlearn}
  Q(\bm{x}_t, \bm{u}_t) = (1-\alpha) Q(\bm{x}_t, \bm{u}_t) + \alpha (R + \gamma Q(\bm{x}_{t+1}, \bm{u}_{t+1}))
\end{equation}
where $Q$ is the expected value of performing action $u \in \bm{u}$ in state $x \in \bm{x}$, $R$ is the reward, $\alpha$ is the learning rate and $\gamma$ is the discounting factor \cite{qlearning}. The learning rate determines the weight given to new information and the discounting factor determines how the emphasis the agent places on future rewards.

\paragraph{Network Exploration}
Reinforcement learning has been used to learn the trustworthiness of target agents based on the strength of trust paths leading into the target agent \cite{KIMSONG11}. This is done by first initializing a trust graph such that each node represents a user and each edge represents the relationship between two users. Each edge is weighted with the direct trust value $\tau$ between the two users.

Two methods for Q-learning were proposed in \cite{KIMSONG11}. For both methods, the agent starts from the source node and selects one of its neighbours using $\varepsilon$-policy.

In the min-max aggregation approach, upon choosing the next node $v_j$ from the current node $u_i$, the agent receives a reward $r_{strength}$
\begin{align}
  &r_{strength}(ui, v_j) \\
  &=
  \begin{cases}
     -1, \qquad & $if $ v_j $ has already been visited$ \\
    \tau(u_i, v_j), \qquad & $otherwise$
  \end{cases}
\end{align}
which rewards the agent based on the trust outcome of the node choice. The $-1$ punishes the agent for visiting existing nodes to avoid the formation of cycles. The agent can then learn from this reward and update the expected return from taking this particular trust edge with
\begin{align} \label{eq: ksrlQstrength}
  &Q^t_{strength}(u_i, v_j) \\ &= (1-\alpha) \cdot Q^{t-1}_{strength} (u_i) + \\
  &\alpha_t \cdot \min \left( r_{strength}(u_i, v_j), Q^{t-1}_{strength}(v_j, v^*) \right)
\end{align}
which defines the terms for Q-learning in Equation \ref{eq: qlearn} for the context of trust. The expected future reward is determined by the node's trustworthiness or the benefit of following the optimal policy for the subsequent action \cite{KIMSONG11}. This learning process is repeated until one of the neighbouring nodes of the target is reached. A path to the target node is now available. Using the learned expected edge rewards, nodes are chosen using the optimal policy
\begin{equation}\label{optimalpolicy}
  \pi^*(u_i) = \argmax_{v_j \in \text{Neighbour}(u_i)} Q_{strength} (u_i, v_j)
\end{equation}
starting from the source node to reach the target node. The trust strength of the target is determined by the trust strength of the path, optimised by reinforcement learning, which is defined to be the minimum value of the trust edge.

A more complex method proposed by \cite{KIMSONG11} for indirect trust computation is the weighted mean aggregation method which attempts to find the path strength to all neighbouring nodes. Here the reward for each node has to redefined to be
\begin{align}
    &r_{path}(u_i, v_j) \\ &=
    \begin{cases}
     -1, \qquad & $if $ v_j $ has already been visited$ \\
     1, \qquad & $if target's neighbour node is reached$ \\
     0, \qquad & $otherwisep$
  \end{cases}
\end{align}
to learn whether or node reaching a particular node would result in a cycle being formed and avoid such a situation. The learning proceeds via the Q-learning equation
\begin{multline}
  Q^t_{path}(u_i, v_j) \\
  = (1-\alpha) Q^{t-1}_{path} (u_i, v_j) + \alpha (r_{path} (u_i, v_j) + \gamma Q^*_{path} (v_j, v*))
\end{multline}
which is similar in the way it employs $Q^*$ value like in Equation \ref{eq: ksrlQstrength} but uses a averaging method instead like that seen in Equation \ref{eq: qlearn}. In the weighted average method, instead of moving on to the next node and repeating the learning process, the agent must first repeat visiting each node from the current node. This time it visits nodes that have a $Q^t_{path} > 0$ to avoid nodes that are likely to form cycles. After doing so, the agent can learn the strength of the trust path using another Q-learning equation
\begin{align}
  &V^t_{strength} \\
  & = (1-\alpha) V^{t-1}_{strength}(u_i) + \\
  &\alpha \left( \frac{\sum_{v_j \in C(u_i)}r_{strength}(u_i, v_j)V^{t-1}_{strength}(v_j)}{\sum_{v_j \in C(u_i)}r_{strength}(u_i, v_j)} \right)
\end{align}
where $C(u_i)$ is the set of neighbours $u_i$ trusts. Here, the immediate reward is defined
\begin{equation}
    r_{strength}(u_i, v_j) = \tau(u_i, v_j)
\end{equation}
until the node $v_j$ is reached. At the target node, the final $r_{strength}$ is defined to be $1$ and $V_{strength}(v_j) = \tau(u_i, v_j)$.

\subsubsection{Artificial Neural Networks}
Artificial neural networks is a machine learning paradigm that centers around mimicking the manner in which the human brain learns. An artificial neural network consists of simple processing units called \textit{neurons} and \textit{weighted connections} between those neurons. In neural networks, there may be multiple layers of neurons, such as the input layer, hidden layer and output layer, that are connected to each other and within the layer depending on the type of neural network. Between two neurons $(i, j)$, the weight is defined by a function $\omega((i,j))$ \cite{neuralnetworktheory}. Each neuron receives either inputs to the network or the output of other neurons and processes these as inputs to the \textit{propagation function} \cite{neuralnetworktheory}. The output of the propagation function and the previous \textit{activation state} of the neuron acts as input to the \textit{activation function} which will output the new activation state of the neuron \cite{neuralnetworktheory}. Finally, the activation acts as input to the \textit{output function} which gives the data output for other neurons \cite{neuralnetworktheory}.

\paragraph{Basic neural network trust model}
The most basic neural network has been used for trust modelling in \cite{ZOLFAGHAR2011833,DSNN,fuzzynn,vehicle20}. In \cite{ZOLFAGHAR2011833}, different features were modelled using the methods in Section \ref{subsubsec: relationalmeasures} and other heuristics. This generated a set of features which then underwent exponential time-based weighting to reduce the weight that the older feature values have. These feature values were then used as input to a neural network forming the input layer. This neural network then outputs which nodes with no previous trust relationship is likely to have a trust relationship and the strength of this newly formed trust.

In \cite{vehicle20}, the neural network is implemented in context of the vehicular network digital environment to determine if the vehicle is trustworthy based on the receiving and delivery time of the messages by the vehicle. A simple ratio of Euclidean distances which represents the time taken for a particular vehicular node to deliver a particular message it receives was used as a proxy for trust values. Since there may be transmission errors, the computed trust value may not always be consistent with the expected trust values. Therefore, the Euclidean distances are adjusted using a neural network to obtain a more reflective trust value.

The message receiving vehicular nodes was used as the neurons for the input layer to the neural network. These took in the numerator Euclidean distance (representing the time take for the receiving vehicular node to receive message) as inputs. The next layer - the hidden layer - consisted of the message forwarding nodes that take in the denominator Euclidean distance (representing the time taken for the message to be forwarded after it is first delivered by the previous sender). In the output layer, the nodes receive the ratio, which is the trust value. When the trust values that were expected and the actual trust values are inconsistent, back propagation is performed and the Euclidean distances are adjusted by the respective nodes in the hidden and input layer. This way, a more representative trust value is obtained and stored for future interactions.

\paragraph{Fuzzy logic and neural networks} \label{par: fuzzynn}
Neural networks can be used in conjunction with other computational methods. In \cite{fuzzynn}, features obtained directly from data sets of a user were used as input into a fuzzy logic module. In this module, fuzzy logic was used to determine the degree to which each particular feature was considered "low", "average", "high" or "very high". This was done using membership functions for each category which produced the degree of membership for each feature in each of the categories. How this is done is further discussed in Section \ref{subsubsec: fuzzylogic}.

These membership degrees were stored in the neurons in the input layer of the neural network. This neural network fused these inputs with a rule layer which consists of $45$ neurons to cover all the combinations of inputs. The rule layer acts as input to the output layer which only has 4 neurons. Each neuron in the final output layer represents the membership function of the trustworthiness of the user in each of the four categories - "low trustworthiness", "average trustworthiness" and so on. The trustworthiness of the user is the category in which the user has the highest membership degree. Training of this neural network was done with the gradient descent method and back propagation algorithm.

\paragraph{Dempster Shaffer theory and neural networks} \label{par: dsnn}
In \cite{DSNN}, features were constructed using functions from Section \ref{subsubsec: relationalmeasures} and heuristics about a user. These features were used as input to a Dempster Shaffer Theory module. In the module, mass functions were assigned to each feature to obtain an evidence prototype. Dempster's rule of combination was then used to combine the different sources of evidence to derive a joint mass function for each feature value. More details of Dempster Shaffer Theory will be discussed in Section \ref{subsubsec: dstheory}.

The joint mass function for each feature value formed a single neuron in the input layer of the neural network. In the next layer, the local fusing layer, the neurons are trained by the data set based on the joint mass functions from the inputs. The best outputs from the local fusing layer form the masses which are trained in the neurons in the global fusing layer. In each node of the fusing layer, a logistic sigmoid activation function was used in each node. Finally, the output layer gives the mass functions for the event that the user is trustworthy and the event that the user is untrustworthy. The trustworthiness is thus determined by the most likely trustworthiness event. This neural network was also learned using the standard back propagation with gradient descent approach.

\paragraph{Bernoulli Neural Network}
Besides the most basic neural network, more complicated constructions have been implemented in trust management. One such example is the Bernoulli neural network implemented by \cite{COBRA}. In the Bernoulli neural network, there are three layers - the input, hidden and output layer. The defining characteristic of the Bernoulli neural networks is that the hidden layer uses Bernoulli polynomials as activation functions \cite{bnntheory}. The $n-1$-th hidden layer neuron can be computed recursively with
\begin{equation}
  \phi_{n-1} (x) = x^{n-1} - \sum^{n-2}_{k=0} \binom{n}{k} \phi_k(x)/n
\end{equation}
which is the recursive form of the Bernoulli polynomial. The input and output layer each have one neuron, both activated by a simple linear function $f(x) = x$. The weight between the input neuron and the hidden layer neurons is set to be 1. The weights between the hidden layer and the output layer neuron is set to be $\omega_j$ where $j=0,1, \dots, n-1$ which should be decided or adjusted. Let the input into the network be $x$ and the output be $y$. From the structure of the entire neural network, the output of the network is
\begin{equation}
  y = \omega_0 \phi_0(x) + \omega_1 \phi_1(x) + \dots + \omega_{n-1} \phi_{n-1}(x).
\end{equation}

A Bernoulli neural network is implemented as part of the trust model in \cite{COBRA}. In the model, advisor agents train models individually, using contextual features, based on past interactions to output a predicted conditional probability about the trustworthiness of the trustee. This recommendation acts as evidence which is aggregated together with the truster's own first-hand evidence using the Bernoulli neural network. The neural network trains its weights using gradient descent back propagation with a cross-entropy loss function.

\paragraph{Deep Belief Neural Network}
Deep Belief Neural Networks(DBN) are neural networks with many hidden layers to perform a deep hierarchical representation of the input data \cite{dbnintro}. One type of DBN uses the Restricted Boltzmann machines(RBM) in each layer of the neural network \cite{dbnintro,dbndeeplearning}. RBMs is a stochastic network that consists of two layers of nodes - a hidden layer and a visible layer \cite{dbndeeplearning}. Each node in one layer is connected to all the nodes in the other layer with weights and vice versa so that the values in one layer affects the values of the other. Nodes take on a value of either $0$ or $1$ at different time intervals with probability conditioned on the nodes in the other layer \cite{dbndeeplearning}. RBM learns by unsupervised learning by presenting training patterns to the visible nodes \cite{dbnintro}. The weights of connections and the biases in each layer is adjusted to minimize the energy of the network \cite{dbnintro}.

DBN uses RBMs in each of its layers by constructing the neural network such that top hidden layer of one RBM acts as visible bottom layer of the RBM layer above it \cite{dbndeeplearning}. Training of the DBN is done by first iteratively training each RBM layer using unsupervised learning to obtain ideal parameters for extracting features from the data \cite{dbndeeplearning}. Then supervised learning classification using methods such as back propagation with gradient descent is performed to fine tune the weights throughout the whole network \cite{dbndeeplearning}.

This DBN training method has been used in trust modelling to achieve context aware trust values \cite{DBNcontextaware}. In \cite{DBNcontextaware}, the DBN model is trained to link the features of a situation - features of the trustor, trustee and context of interaction - with trust values. When there is insufficient evidence for a specific interaction - the specific trustor, trustee and context features - the DBN neural network can still derive a trust value even with little information by using incompletely related past interactions.

\cite{DBNcontextaware} achieved this using a DBN with an input layer, three hidden layers and a label layer. The input layer takes in normalized feature values and passes them into the first hidden layer. The hierarchical nature of DBN is used here to progressively filter more important features at each layer. At the first hidden layer, there are nodes corresponding to the total number of features. At the second hidden layer, only more significant features are selected by restricting the total number of nodes to a fraction of the total number of features. At the last hidden layer, the nodes are fused to decide a trust value for the specific context. The final label layer outputs a single value representing the trust value for the current interaction.

\paragraph{Growing Hierarchical Self-Organising Map}
Self-organising maps(SOM) are a unsupervised learning, neural network model that preserves the topological in the input space into the output space \cite{topologypreservation, somtheory}. This topological preservation means that the similarity of the input data is mirrored to a very large extent in geographical vicinity within the representation space \cite{ghsomtheory}. In SOMs, neurons are organised in a two-dimensional rectangular or hexagonal grid and each neuron is assigned a weight vector of the same dimension as the input vector \cite{somtheory}. At each iteration, SOM finds the weight vector that is closest to the input vector in the data \cite{somtheory}. The SOM algorithm updates the weight of this vector and that of its surrounding nodes while maintaining the connections from the original grid \cite{somtheory}. This is repeated until the map converges

Growing Hierarchical Self-Organising Map (GH-SOM) is a variant of SOM that independently determines the topological space (which needs to be decided prior to training in SOM) and mirrors the hierarchical relations in the data \cite{ghsomtheory}. This is done by a hierarchical structure of multiple layers where each layer is several independent SOMs. The first layer contains one SOM and for every neuron in the SOM, an SOM can be added to form the next layer. This expansion helps represent the subset of data at the specified level of granularity \cite{Capua}.

Capua utilised GH-SOM to classify content on social networks as harassment or non-harassment. By collecting different features about the content, \cite{Capua} collected the features into a an input vector. These input vectors representing features of a specific content piece are used as input data to a GH-SOM. The GH-SOM then groups in a hierarchical manner, independently, the different content based on their features.

\section{Future Work and Conclusion} \label{sec: futurework}
In our survey paper, we analysed a broad range of environments in the digital world. Then using our understanding of the digital world, we constructed a suitable understanding of trust as a concept and as a soft security mechanism. Finally, we explained different trust modelling methods and how they have been used so that their theoretical basis and usefulness can be understood. In writing this survey, we hope to have offered a well-rounded survey of trust management from analysis of application environment to actual modelling methodology.

Naturally, there are a number of areas in which our survey can improve on. For one, we did not manage to align our digital environment analysis and factors with modelling methodology. An additional step in our survey would have been to tie in which factors have been modelled and how they can be modelled using existing mathematical methods. However, we were not able to do so as the scope of the survey would have been too big.

Another aspect of trust that we could have explored was the suitability of each method for security. While we covered the theoretical basis of each model, we did not manage to perform any analysis on the suitability of each method and how each method would evaluate trust in the face of security attacks. Combinations of methods --- whether they complement or conflict each other --- were also not considered . Further analysis can be performed in the future to determine how sensitive each method is to different trends and behaviour in the system. This pushes research in the direction of the most suitable technical method for trust management.

Lastly, the the lack of real-world data about trust makes research in this field particularly challenging. While some real world data is available about some digital environments, it is rare to have actual data with corresponding trust evaluation, much less data sets that are recent. Methods to expand on data sets or to make use of existing data sets in meaningful ways should be analysed. Alternatively, using real world data, sample and simulation data can be created. This was not considered or used for evaluation in our survey.

Finally, from our derivation of different factors based on the digital world, we find that there are many tangentially related fields that can aid trust management. Future surveys in trust can consider looking into behavioural management, anomaly-detection and risk management models to see how, when abstracted, they can function as trust management mechanisms. This could be a good direction for future trust model construction.

\section*{Acknowledgment}

We thank the editor and anonymous referees of this journal whose comments substantially improved this article.

\bibliographystyle{IEEEtran}

\EOD

\end{document}